\documentclass[doublecol]{epl2}

\usepackage{amsmath}
\usepackage{amsthm}
\usepackage{lineno}

  \title{Dynamics of a ring of three unidirectionally coupled Duffing oscillators with time-dependent damping}

  \shorttitle{Dynamics of a ring of Duffing oscillators with time-dependent damping} %Insert here a short version of the title if it exceeds 70 characters

  \author{J. J. Barba-Franco\inst{1} \and A. Gallegos\inst{1} \and R. Jaimes-Re\'ategui\inst{2} \and S. A. Gerasimova \inst{2} \and A. N. Pisarchik \inst{2,3,4}}
  \shortauthor{J. J. Barba-Franco \etal}

  \institute{                    
  \inst{1} Departamento de Ciencias Exactas y Tecnolog\'{\i}a, Centro Universitario de los Lagos, Universidad de Guadalajara, Enrique D\'{\i}az de Le\'{o}n 1144, Colonia Paseos de la Monta\~{n}a, Lagos de Moreno, Jalisco, Mexico\\
  \inst{2} Lobachevsky State University of Nizhny Novgorod, Nizhny Novgorod, Russia\\
  \inst{3} Centro de Tecnolog\'ia Biom\'edica, Universidad Polit\'ecnica de Madrid, Campus de Montegancedo, Pozuelo de
  Alarc\'on, 28223 Madrid, Spains\\
  \inst{4} Innopolis University, Universitetskaya Str. 1, 420500 Innopolis, Republic of Tatarstan, Russia
}


\abstract{
We study dynamics of a ring of three unidirectionally coupled double-well Duffing oscillators for three different values of the damping coefficient: fixed dumping, proportional to time, and inversely proportional to time. The dynamics in all cases is analyzed through time series, Fourier and Hilbert transforms, Poincar\'e sections, as well as bifurcation diagrams and Lyapunov exponents with respect to the coupling strength. In the first case, we observe a well-known route from a stable steady state to hyperchaos through Hopf bifurcation and a series of torus bifurcations, as the coupling strength is increased. In the second case, the system is highly dissipative and converges into one of stable equilibria. Finally, in the third case, transient toroidal hyperchaos takes place.}

\begin{document}

\maketitle

\section{Introduction\label{S:Intro}}
Duffing equations describe dynamics of a systems with cubic nonlinearity, which can have either single-well or double-well potential. As distinct from a harmonic oscillator described by a linear second-order differential equation, the Duffing oscillator in its original form has essentially only one extra nonlinear stiffness term. Despite its enigmatic simplicity, the Duffing oscillator was successfully used to model various physical processes such as stiffening strings, beam buckling, superconducting Josephson parametric amplifiers, ionization waves in plasmas, as well as biomedical processes (see~\cite{lakshmanan1996chaos} and references therein). The Duffing equations are easily implemented in electronic circuits. Several physical realizations of the Duffing equation in laboratory-based experimental models are described in the Virgin's book~\cite{virgin2000introduction} and other review papers (see, e.g.,~\cite{virgin2007vibration,kovacic2011duffing} and references therein). In particular, a double-well Duffing oscillator was widely used to evaluate a large variety of nonlinear systems including slender aerostructures that may buckle under loads~\cite{virgin2007vibration}, microelectromechanical switches~\cite{qiu2004curved}, vibration-based energy harvesters~\cite{kazmierski2014energy,harne2013review}, electrical circuits~\cite{debnath1989remarks}, and optical systems~\cite{dykman1991stochastic}. 

Coupled Duffing oscillators have attracted special attention due to their intriguing synchronization behavior~\cite{boccaletti2018synchronization}, including two-state intermittency~\cite{jaimes2004}, transition to hyperchaos~\cite{kapitaniak1993}, intermittent lag synchronization~\cite{pisarchik2005}, attractor annihilation in stochastic resonance \cite{pisarchik2014control}, etc. The coupled Duffing oscillators were extensively studied with respect to their parameters, such as the coupling strength, nonlinearity stiffness term, external force or modulation of parameters accessible to the system have been developed. In particular, a ring of unidirectionally coupled Duffing oscillators exhibit the most interesting dynamics, such as a transition from periodic to chaotic and hyperchaotic behavior and so-called rotating wave~\cite{perlikowski2010routes, borkowski2015experimental, borkowski2020stability}. The stability of this system was estimated using Lyapunov exponents~ \cite{dabrowski2012estimation,balcerzak2018fastest}. Nevertheless, little attention was paid to a study of the effect of the damping term, although ring-coupled overdamped Duffing oscillators were investigated in the presence a delay in coupling~\cite{tchakui2016dynamics}, multistability~\cite{meena2020resilience,jaimesself}, and even proposed for spectrum-sensing technology application~\cite{tang2016rf}. 

It is well known when a motion takes place in the environment, the system has dissipation which is modeled by a damping term related to the velocity \cite{Landau}. In this regard, several papers were devoted to a study of Duffing oscillators with linear and nonlinear damping terms (see~\cite{kovacic2011duffing} and references therein). The knowledge of the potential can be useful not only for conservative quantum systems, but also for understanding dynamics of dissipative systems. Although dissipative systems cannot be described by a proper potential~\cite{graham1984existence}, however, in some cases the potential can still be found. For example, in the case of a linear time-dependent damping term, the system can be viewed as an undamped oscillator but with a variable mass and therefore the corresponding potential can be obtained~\cite{cieslinski2010direct,barba2020lagrangians}. In addition, analytical studies of transitions which occur between three possible dynamical states (cluster synchronization, complete synchronization, and instability) were performed in a ring of $N$ diffusely coupled Duffing oscillator~\cite{kouomou2003transitions, yolong2006synchronization}. 
  
In this Letter, we study the dynamics of three double-well Duffing oscillators coupled in a cyclic ring. As distinct for previous studies of the same coupling configuration, in our work we focus on the effects of the damping term. We consider three different cases of the damping coefficient: fixed damping, damping proportional to time (overdamping case), and damping inversely proportional to time (quasi-conservative case). To characterize the system dynamics we use time series, power spectra, Poincar\'e sections, bifurcation diagrams, and Lyapunov exponents. The third case is the most interesting, because we observe, for the first time to the best of our knowledge, \emph{transient toroidal hyperchaos}.

The rest of the Letter is organized as follows. First, we describe the model of three unidirectionally ring-coupled damped Duffing oscillators. Then, we consider the case of the fixed damping term and demonstrate the existence of the rotating wave in the ring. After that, we study the case when the damping coefficient if proportional to time and demonstrate the overdamped  dynamics leading to a stable fixed point. Finally, we analyze the system dynamics when the damping term is inversely proportional to time and demonstrate transient toroidal hiperchaos. At the end of the Latter, we summarize the main results. 

\section{Model}
Let us consider the simplest form of the undamped Duffing oscillator without a driving force, which can be given as
\begin{equation}
\label{DO1}
\ddot{x} + \omega_0^2 x +\delta x^3 = 0,
\end{equation}
where the point means time derivative and $\omega_0^2$ and $\delta$ are non null real constants. It is easy to find the corresponding potential associated with the motion eq.~(\ref{DO1}), which has the following form
\begin{equation}
\label{POT}
V(x)=\frac{1}{2} \omega_0^2 x^2 + \frac{1}{4} \delta x^4.
\end{equation}
The shape of the potential function eq.~(\ref{POT}) depends on the values of the parameters $\omega_0^2$ and $\delta$. Basically, there are four different cases:
\begin{itemize}
	\item If $\omega_0^2 > 0$ and $\delta < 0$ then the potential has a double-hump well with a local minimum at $x=0$ and two maxima at $\pm \sqrt{\omega_0^2/\left|\delta\right|}$.
	\item If $\omega_0^2 > 0$ and $\delta > 0$ then the potential has a single well with a local minimum at $x=0$.
	\item If $\omega_0^2 < 0$ and $\delta < 0$ then the potential has a single hump with a local maximum at $x=0$.
	\item If $\omega_0^2 < 0$ and $\delta > 0$ then the potential has a double well with two minima at $\pm \sqrt{\left|\omega_0^2\right|/\delta}$ and a local maximum at $x=0$.
\end{itemize}
Here, we are interested in the last case, i.e. bistability. Since a damping term is included in eq.~(\ref{DO1}), the expected motion is ruled by the potential eq.~(\ref{POT}) but with dissipation until the trajectory is attracted to a stable fixed point. The question is if it is possible to define a proper potential for this kind of damped systems. 

It is well known that in general the damped systems do not always have a defined potential~\cite{graham1984existence}. However, it is possible to find potentials for dissipative systems if the damping coefficient can be written as logarithmic derivative of certain function (see, e.g.,~\cite{barba2020lagrangians,cieslinski2010direct}). More explicitly, if we have a motion equation of the type
\begin{equation}
  \label{ODE2}
  \ddot{x}+\alpha(t)\dot{x}+\beta(x,t)=0,
\end{equation}
it can be described by a Langrangian of the form
\begin{equation}
  \label{L4}
  L=\frac{1}{2}m(t)\dot{x}^2-V(x,t),
\end{equation}
where 
\begin{equation}
  \alpha(t)=\dot{m}/m, \qquad V(x,t)=m(t)\int^x \beta(z,t)dz,
\end{equation}
and $\alpha(t)$ is a time-dependent damping coefficient.

In other words, eq.~(\ref{ODE2}) can be viewed as an undamped motion with variable mass $m(t)$. 
Therefore, including $\alpha(t)$ to eq.~(\ref{DO1}) we get
\begin{equation}
\label{DO2}
\ddot{x} + \alpha(t) \dot{x} + \omega_0^2 x +\delta x^3 = 0,
\end{equation}
and the corresponding potential is given as 
\begin{equation}
\label{POT1}
V(x,t) = m(t)\left[ \frac{1}{2} \omega_0^2 x^2 + \frac{1}{4} \delta x^4 \right],
\end{equation}
where $\alpha(t)=\dot{m}/m$. In other words, we are talking about the same potential eq.~(\ref{POT}) with similar fixed points but with a scaling factor $m(t)$. However, we will show in the next sections that different types of time dependence of the damping coefficient can change the dynamics of the ring-coupled Duffing oscillatory system.

The ring of three unidirectionally coupled Duffing oscillators is described as follows~\cite{barba2020lagrangians,jaimesself}
\begin{equation}
\label{ODES1}
  \begin{aligned}
  \ddot{x}_1+\alpha (t)\dot{x}_1+\omega_0^2 x_1+\delta x_1^3+\sigma(x_1-x_3) & = 0, \\
  \ddot{x}_2+\alpha (t)\dot{x}_2+\omega_0^2 x_2+\delta x_2^3+\sigma(x_2-x_1) & = 0,  \\
  \ddot{x}_3+\alpha (t)\dot{x}_3+\omega_0^2 x_3+\delta x_3^3+\sigma(x_3-x_2) & = 0,
  \end{aligned}
\end{equation}
where $\sigma$ is the coupling strength.

In the rest of the paper, we set $\omega_0^2=-0.25$, $\delta=0.5$ which are parameters related to a bistable Duffing oscillator that has three fixed points associated with the potentials eq.~(\ref{POT}) or~(\ref{POT1}) given as  
\begin{equation}
\label{FP}
	\begin{aligned}
	x_{u,d} & = \pm \sqrt{\left|\omega_0^2\right|/\delta} = \pm \sqrt{1/2}, \\
	x_0 & = 0,
	\end{aligned}
\end{equation}
where fixed points $x_{u,d}$ are stable (subindices $u$ and $d$ correspond to positive and values) and $x_0$ is unstable. 

By converting the second-order eq.~(\ref{ODES1}) into first order equations using the change of the variable $\dot{x}=y$, the dynamics of the $j$th oscillator in the ring can be described by the following pair of first order dimensionless ordinary differential equations:
\begin{equation}
\label{ODES9}
  \begin{aligned}
  \dot{x}_j&=y_{j},\\
  \dot{y}_j&=-\alpha (t) y_j+\omega_0^2 x_j-\delta x_j^3+\sigma(x_{j-1}-x_j),
  \end{aligned}
\end{equation}
where $\sigma$ is the coupling coefficient for each oscillator $j=1, 2, 3$. 

In the next sections, we will consider three cases of the time-dependent damping coefficient: $\alpha=0.4$ (constant coefficient), $\alpha=t/4$ (linearly increasing in time), and 
$\alpha= 1/t$ (linearly decreasing in time).

\begin{figure}[th!]
	\centering
        \includegraphics[width=0.45\textwidth]{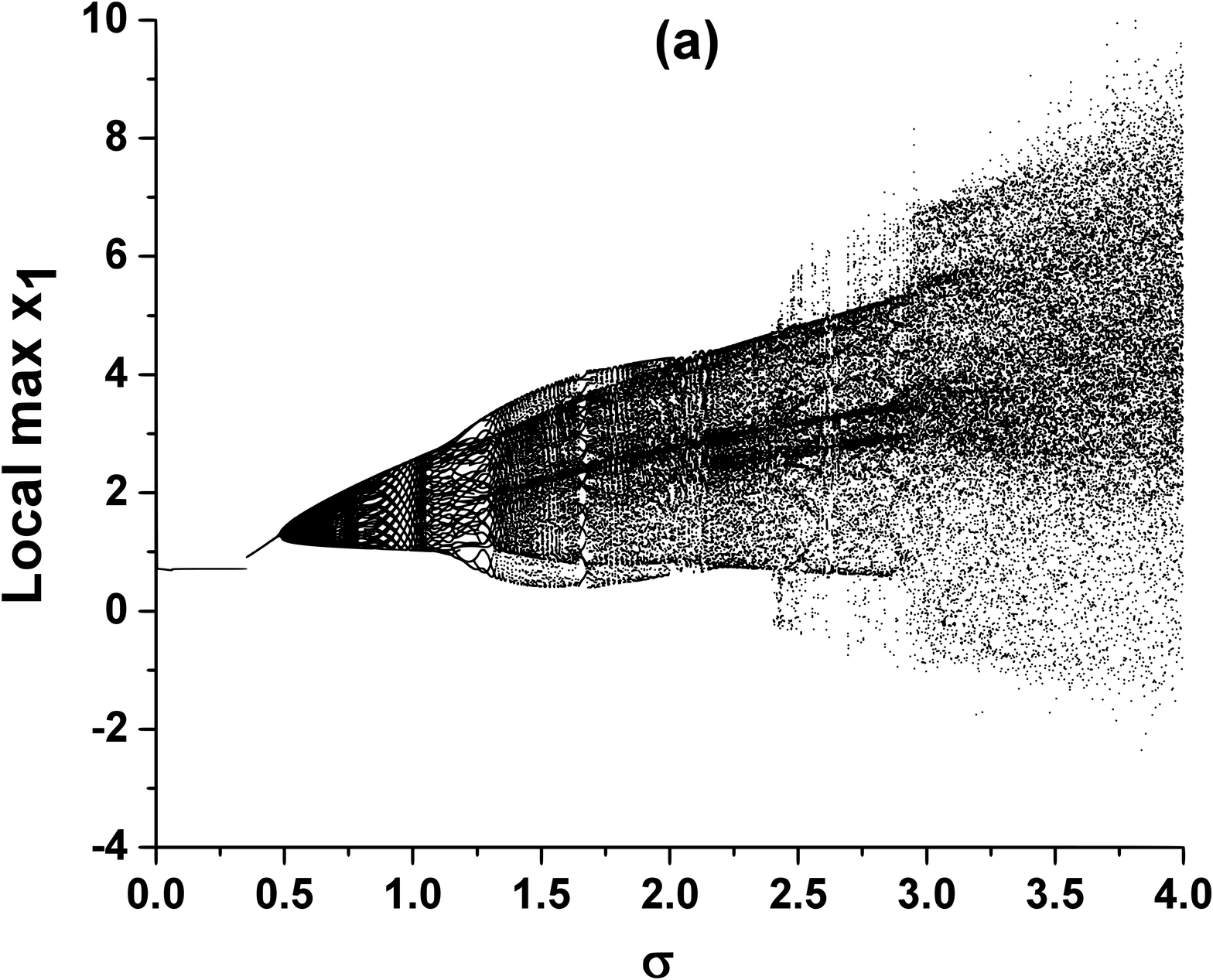} 
        \includegraphics[width=0.45\textwidth]{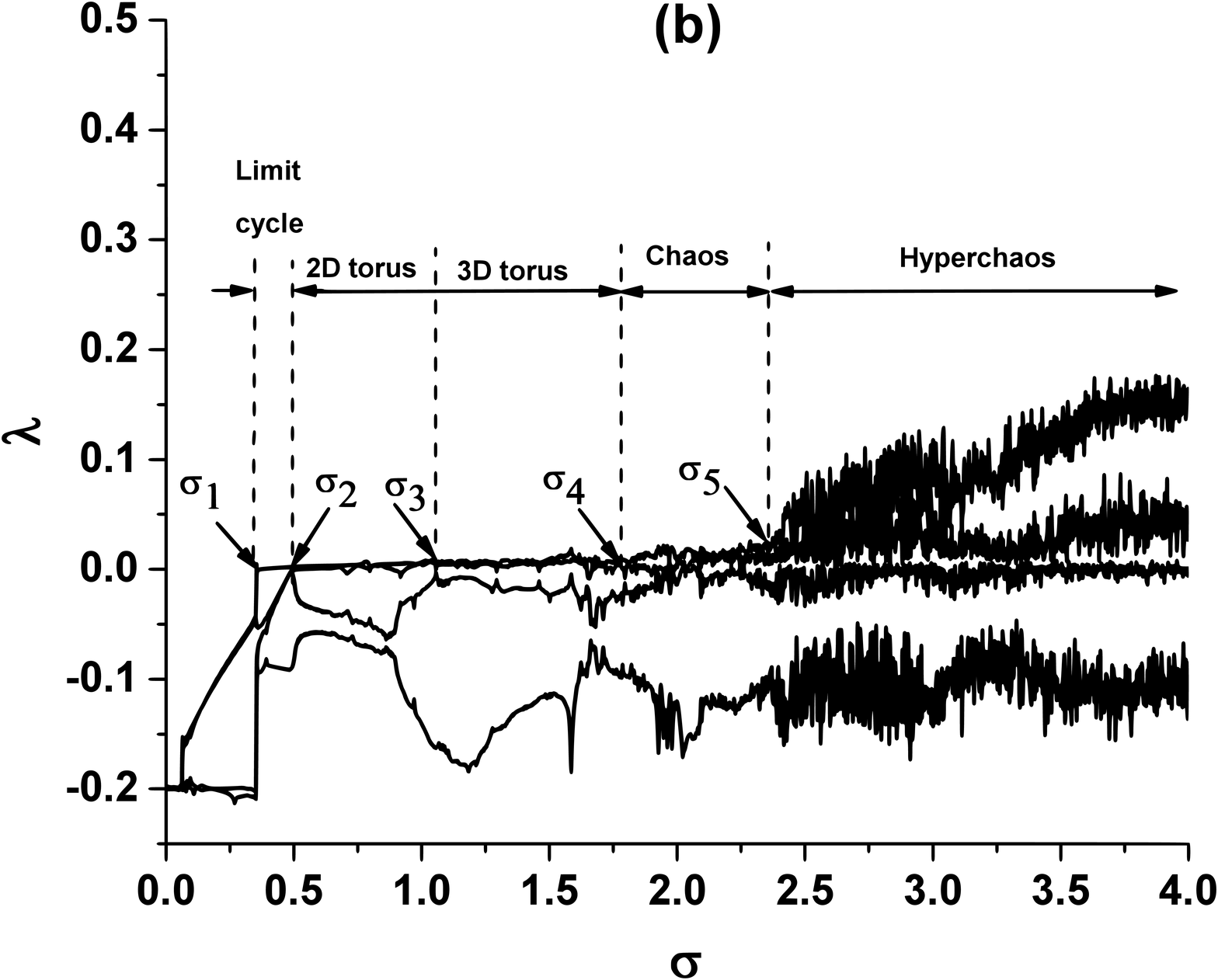} 
        \caption{(a) Bifurcation diagram of the local max of $x_1$ and (b) four largest Lyapunov exponents $\lambda$ versus coupling strength $\sigma$ for $\omega_0^2=-0.25$, $\delta=0.5$, and $\alpha=0.4$.
          \label{fig1}}
\end{figure}

\section{Dynamics of the system with fixed damping}
First, we consider the case of the fixed damping coefficient $\alpha(t)=0.4$. Due to the symmetrical coupling of the three Duffing oscillators eq.~(\ref{ODES1}), the analysis is made with the bifurcation diagram of the local maxima of the amplitude of one of the oscillators (e.g., $x_{1}$) and four largest Lyapunov exponents $\lambda$ with respect to the coupling strength $\sigma$. These diagrams are shown in fig.~\ref{fig1}. The observed  bifurcation scenario from the equilibrium point to chaos and hyperchaos via subsequent Hopf bifurcations is in a good agreement with the Landau-Hopf transition to turbulence~\cite{landau1944problem, hopf1948mathematical} and the Newhouse, Ruelle, and Takens theorem~\cite{newhouse1978occurrence} stated that just after the successive Hopf bifurcation a torus decays into a strange chaotic attractor.

\begin{figure*}[th!]
  \centerline{%
    \begin{tabular}{c}
      (a) \\
      \includegraphics[width=0.27\textwidth]{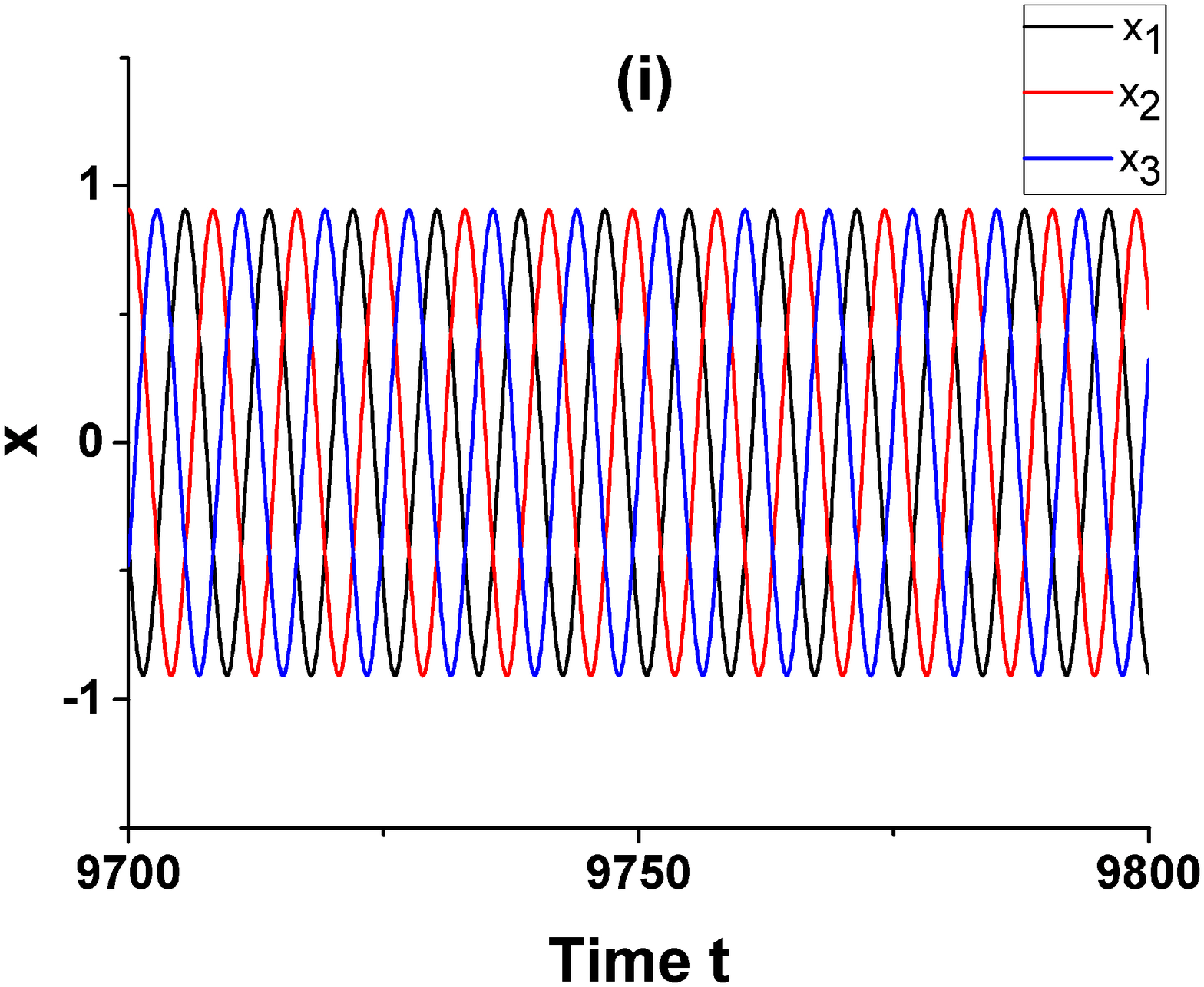} 
      \includegraphics[width=0.27\textwidth]{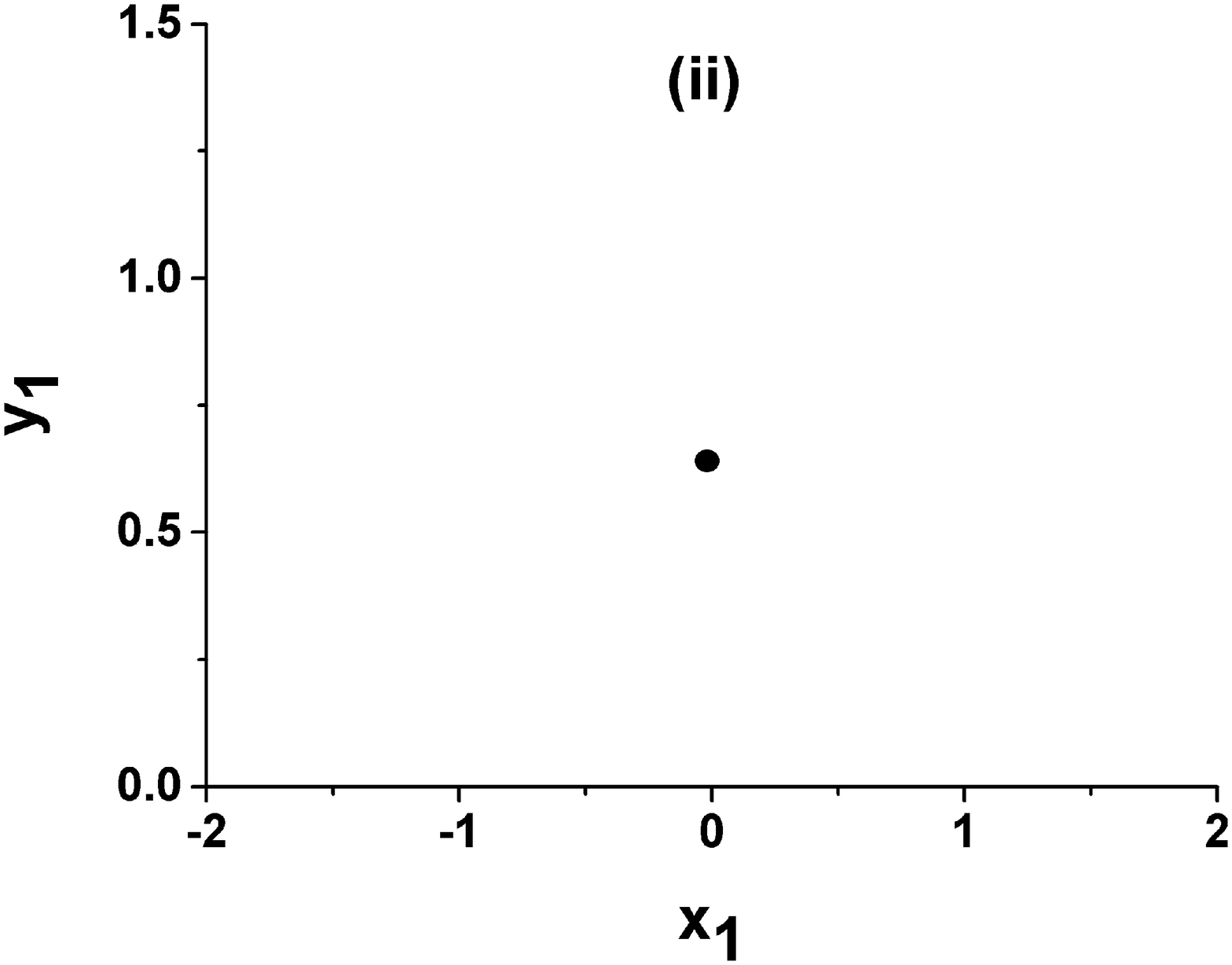} 
      \includegraphics[width=0.27\textwidth]{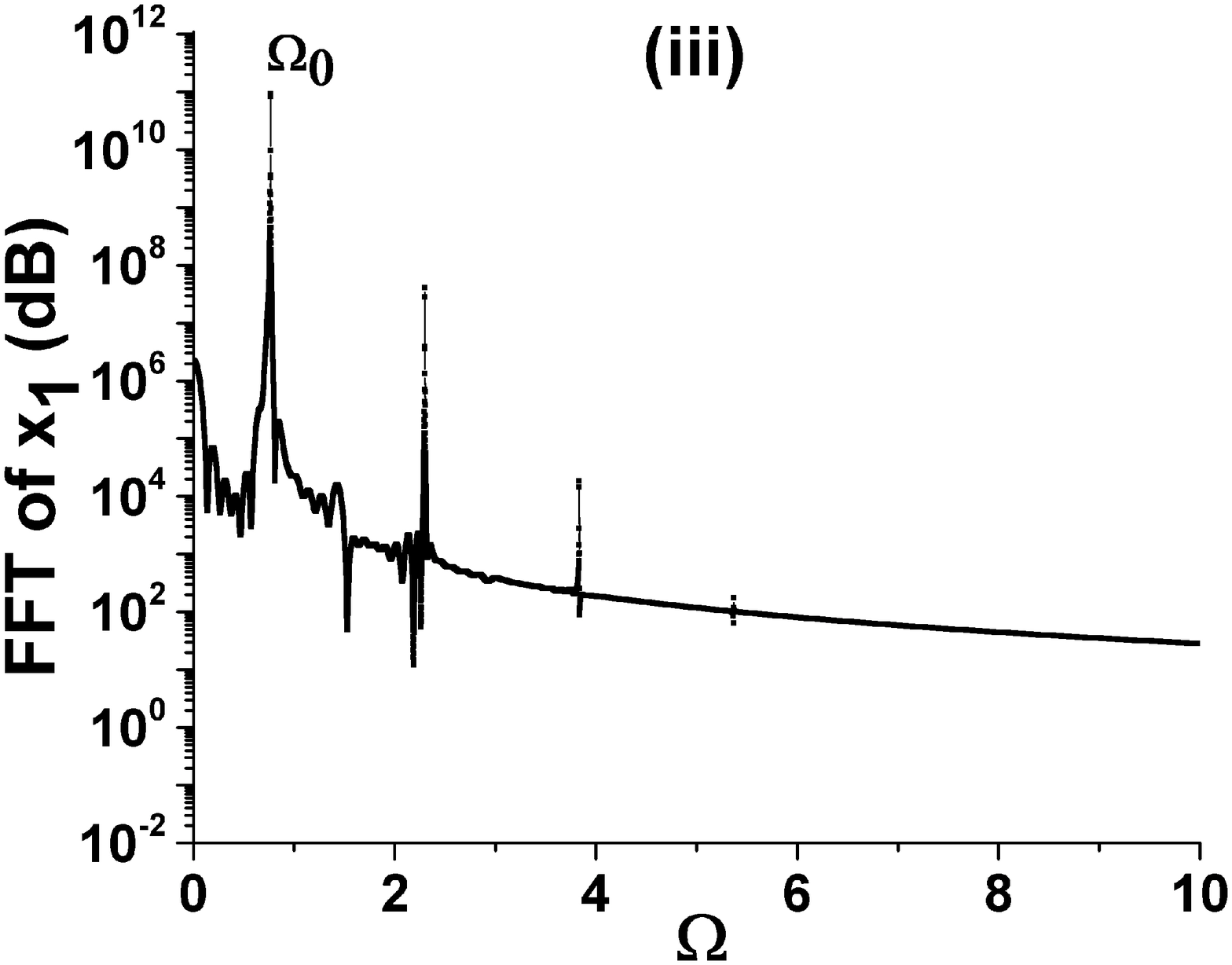}  \\
      (b) \\
      \includegraphics[width=0.27\textwidth]{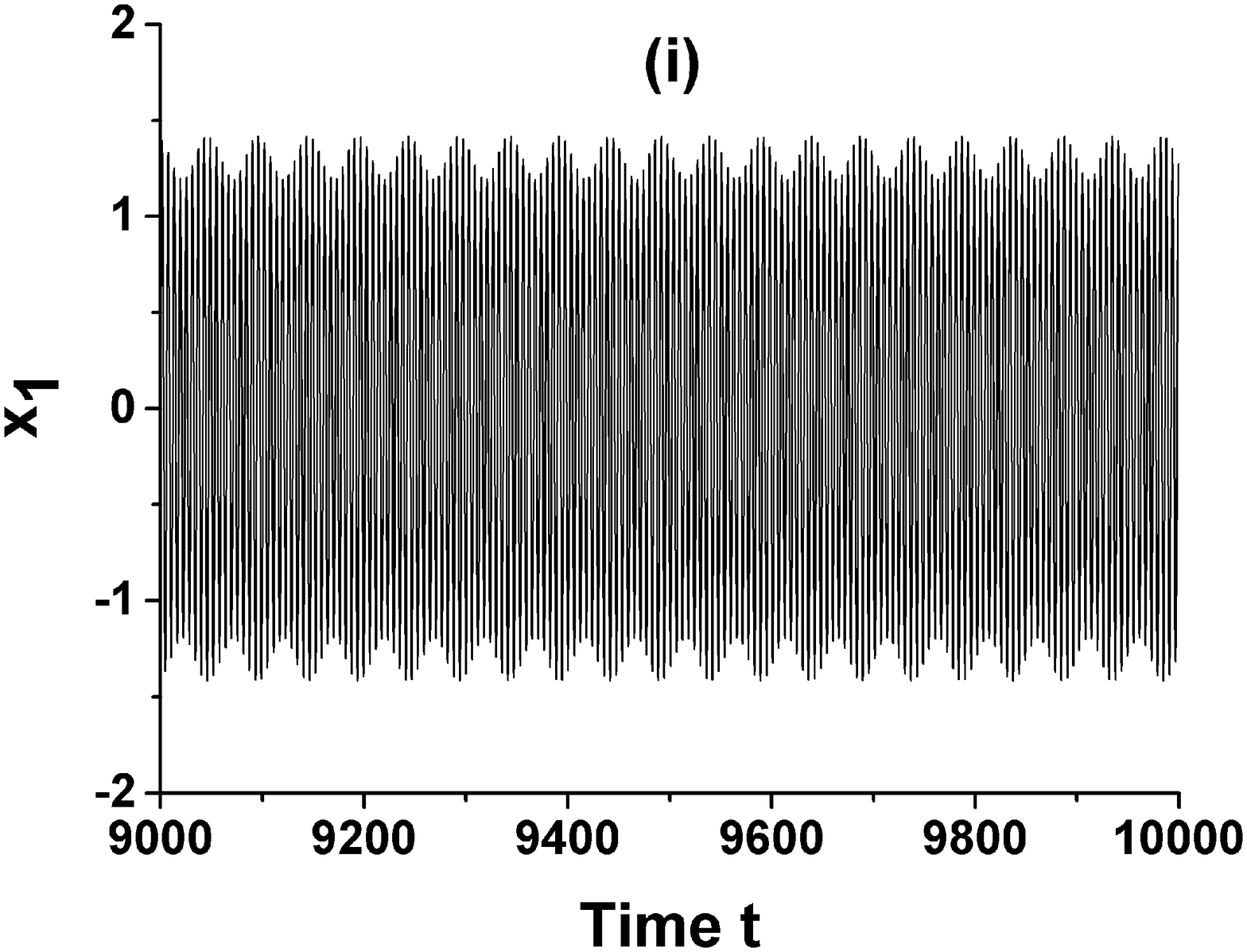} 
      \includegraphics[width=0.27\textwidth]{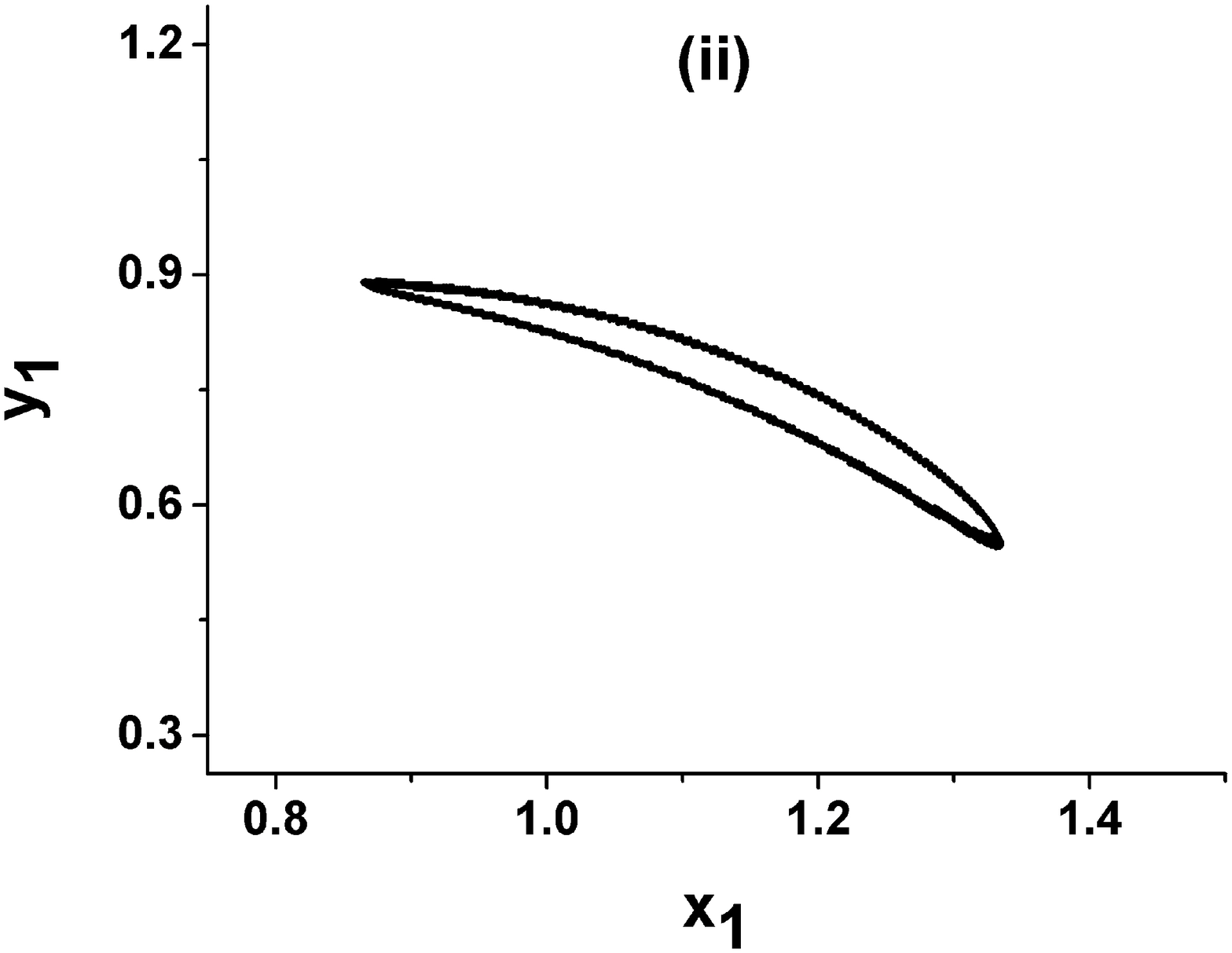} 
      \includegraphics[width=0.27\textwidth]{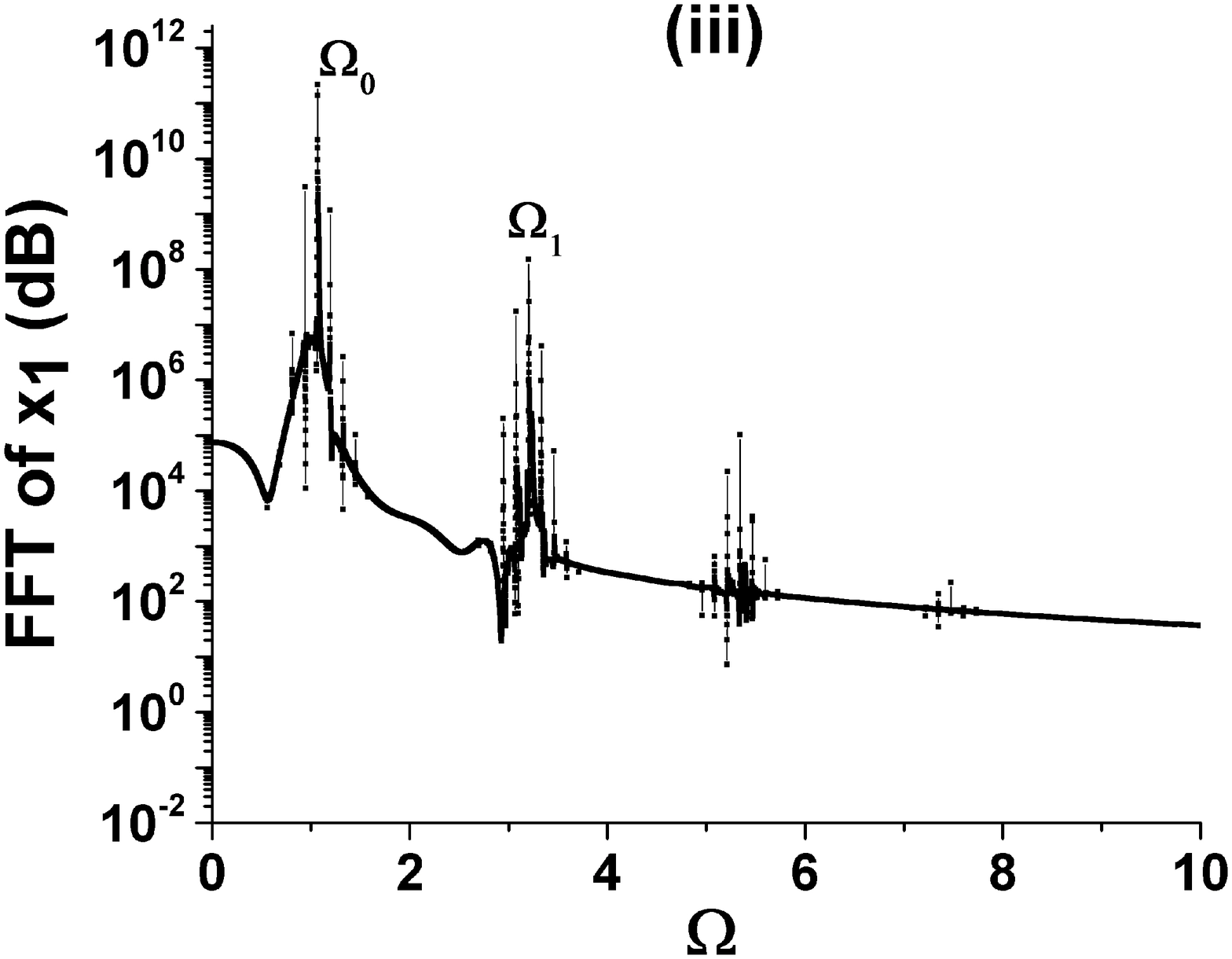}  \\
      (c) \\
      \includegraphics[width=0.27\textwidth]{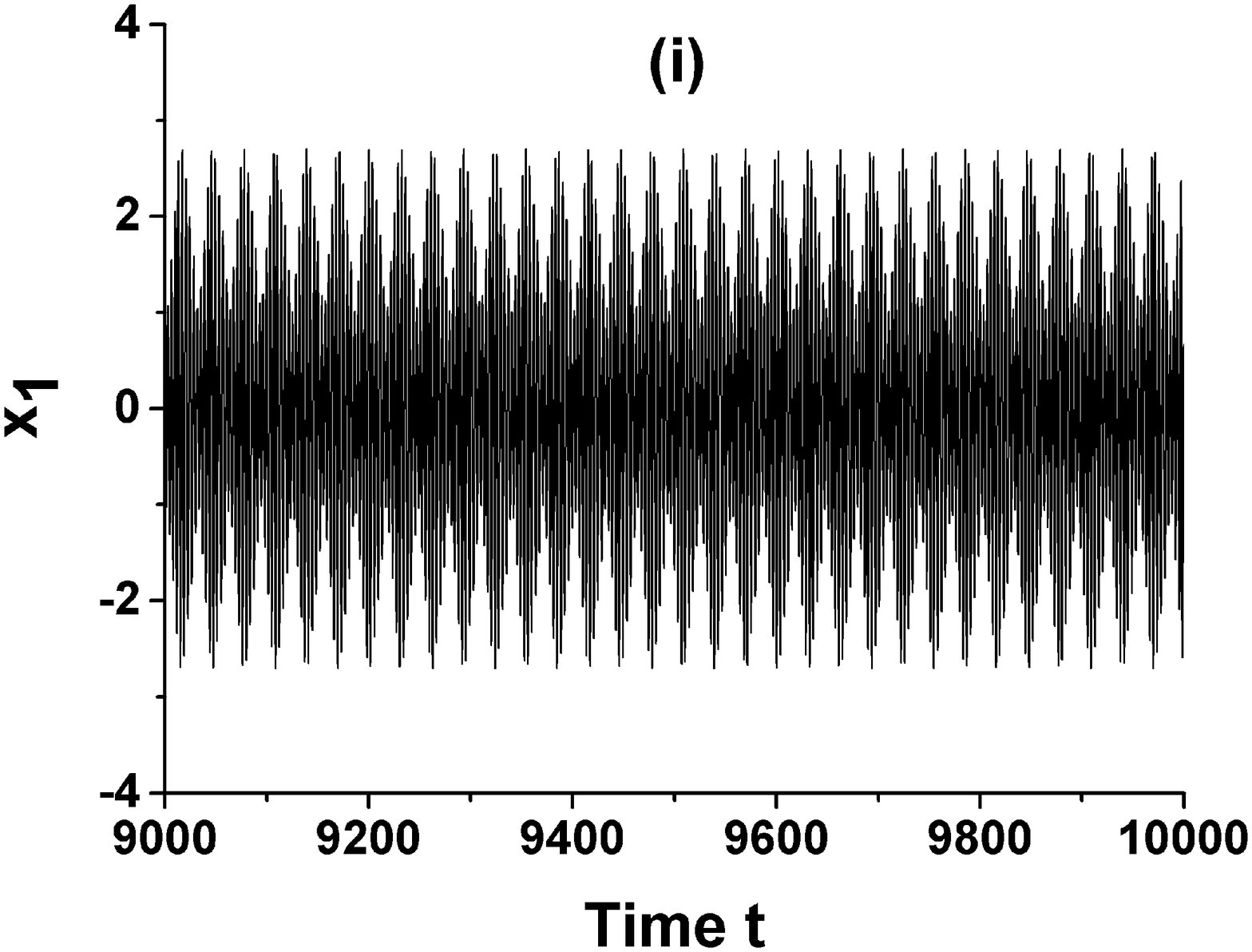} 
      \includegraphics[width=0.27\textwidth]{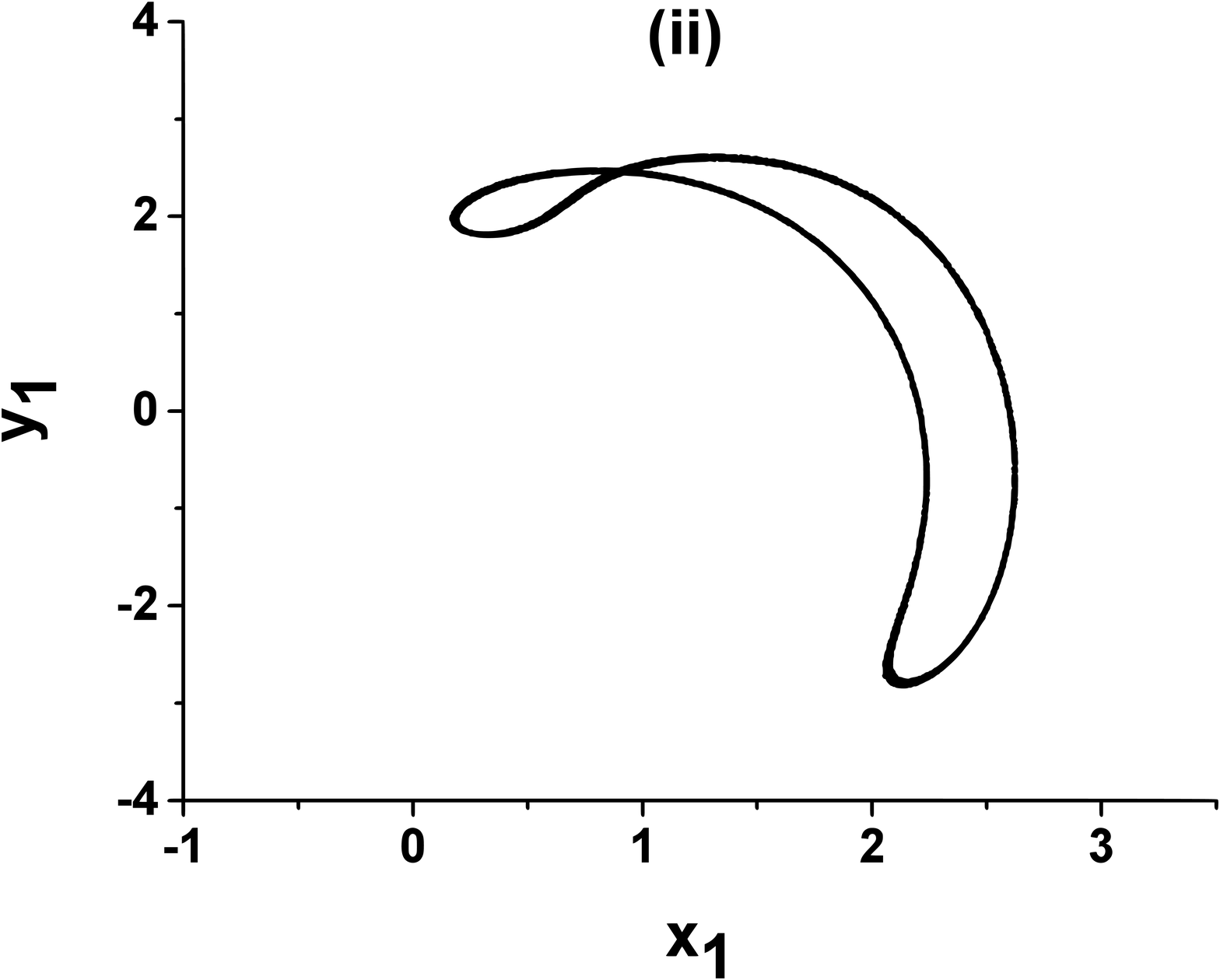} 
      \includegraphics[width=0.27\textwidth]{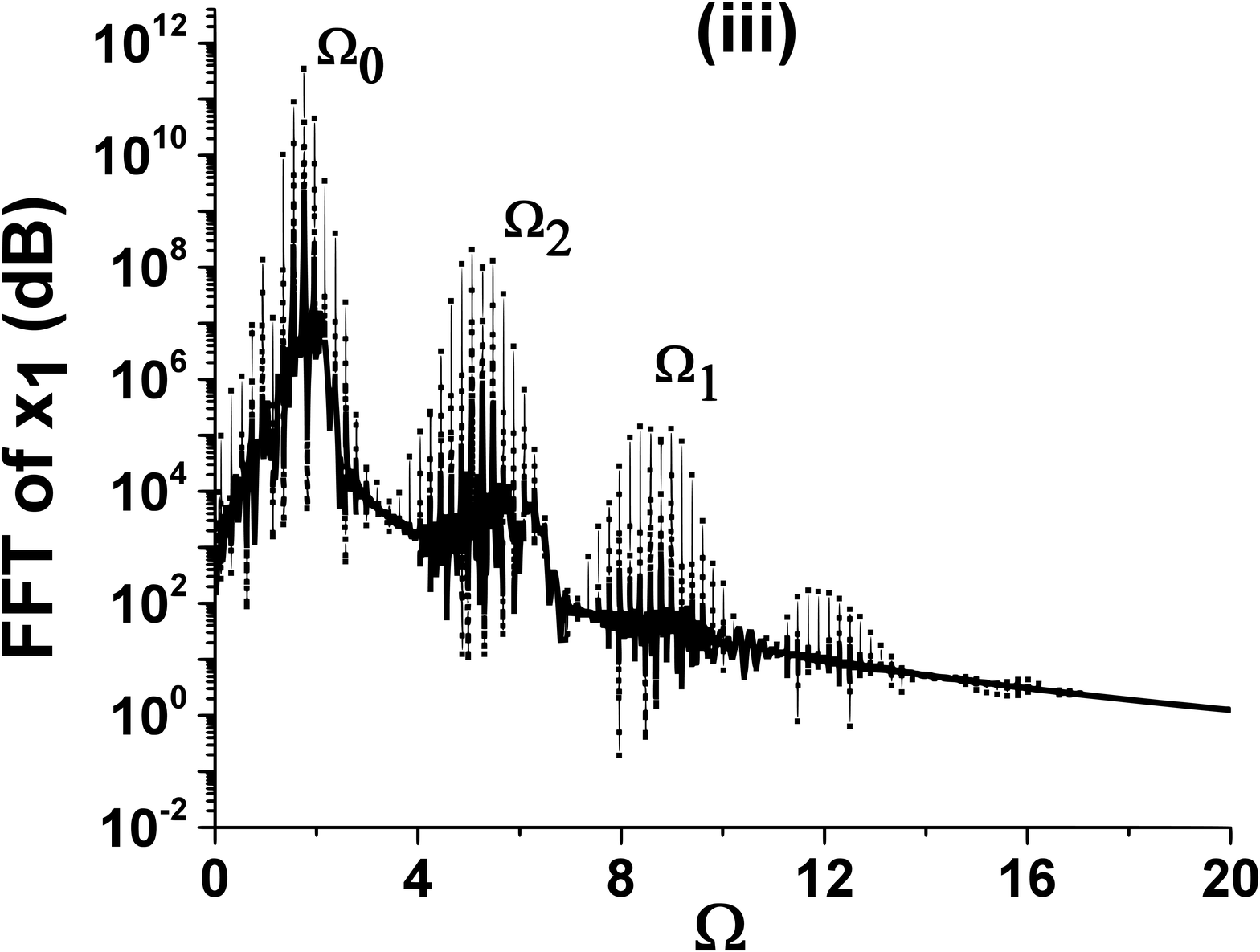}  \\
      (d) \\
      \includegraphics[width=0.27\textwidth]{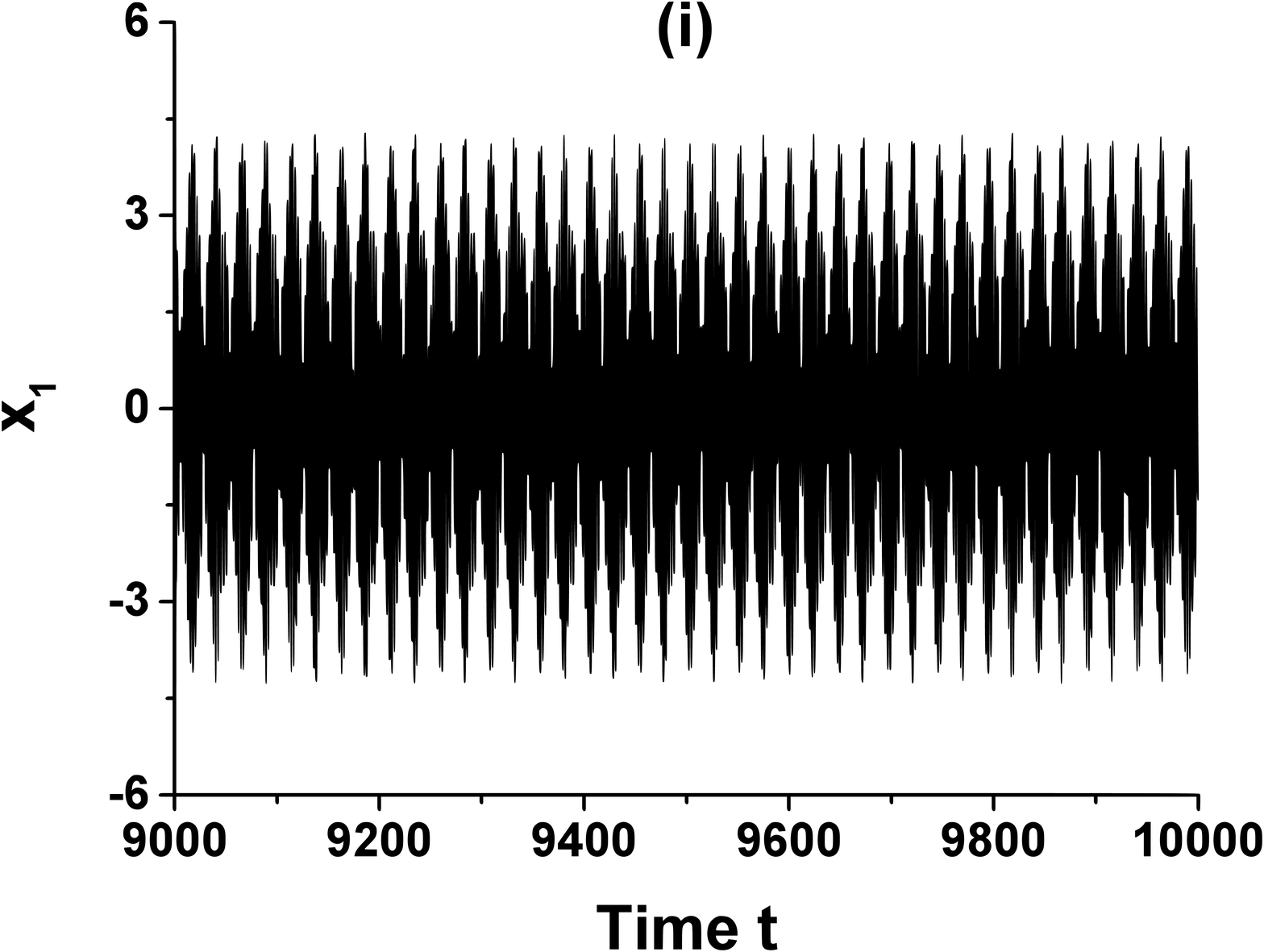} 
      \includegraphics[width=0.27\textwidth]{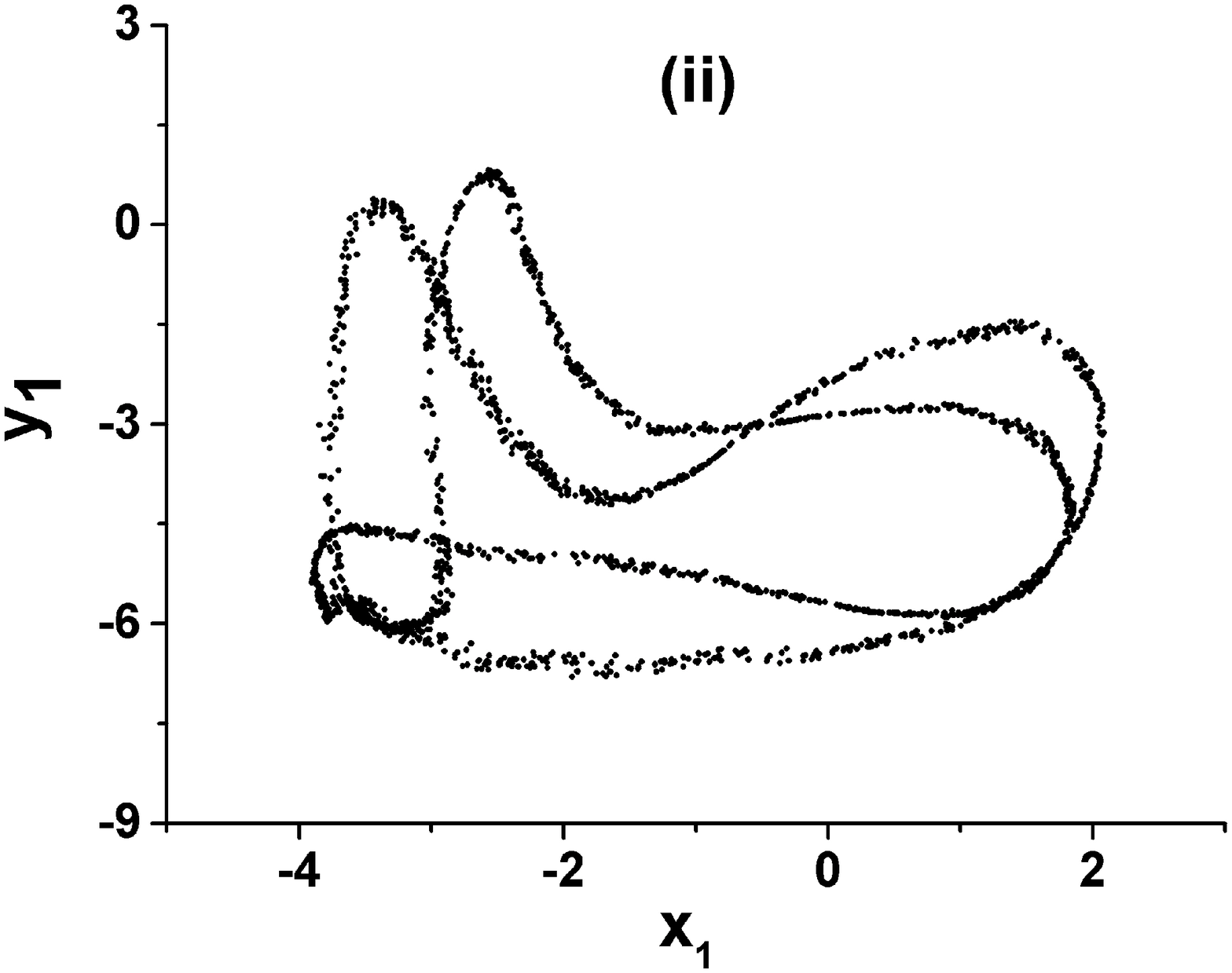} 
      \includegraphics[width=0.27\textwidth]{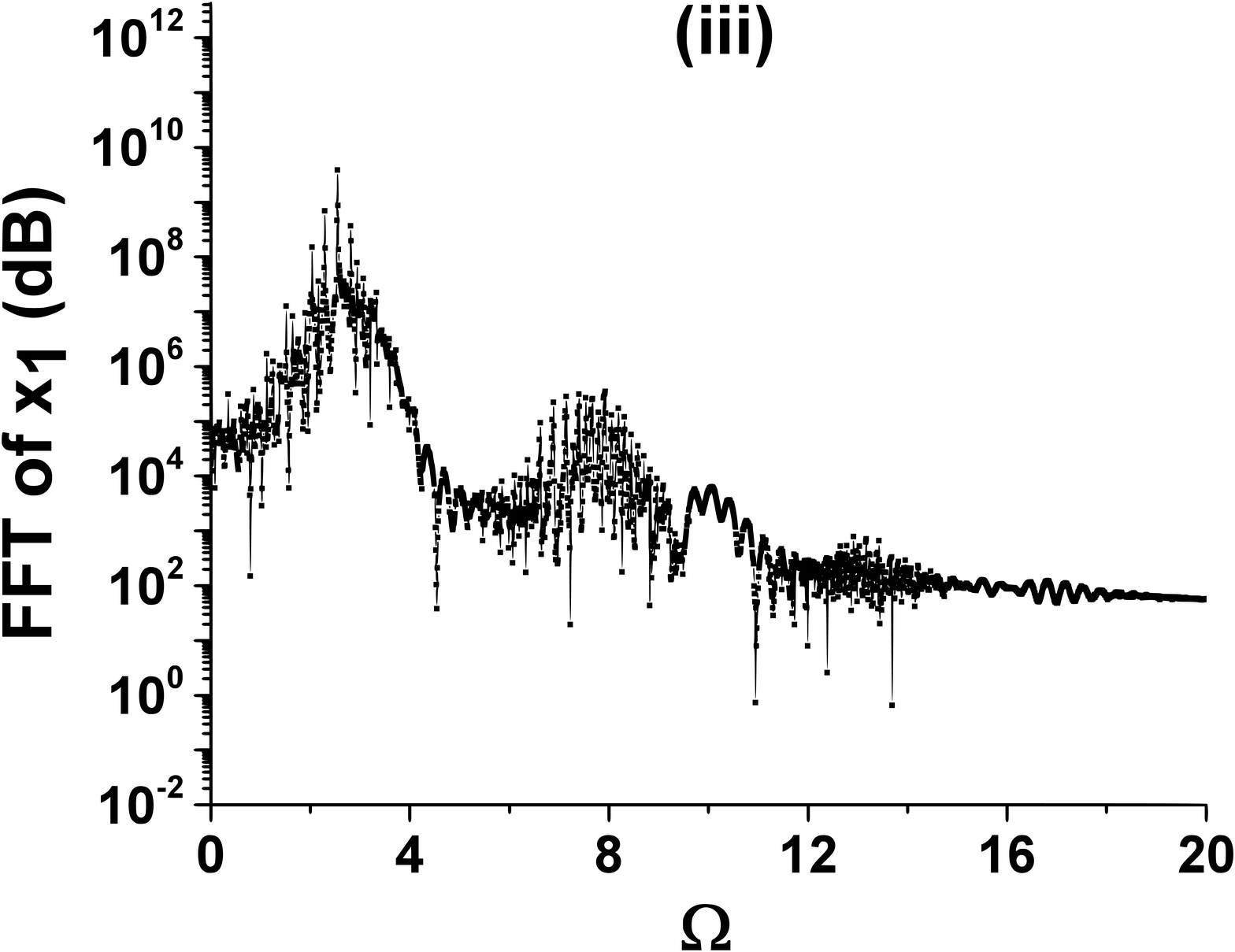}  \\
      (e) \\
      \includegraphics[width=0.25\textwidth]{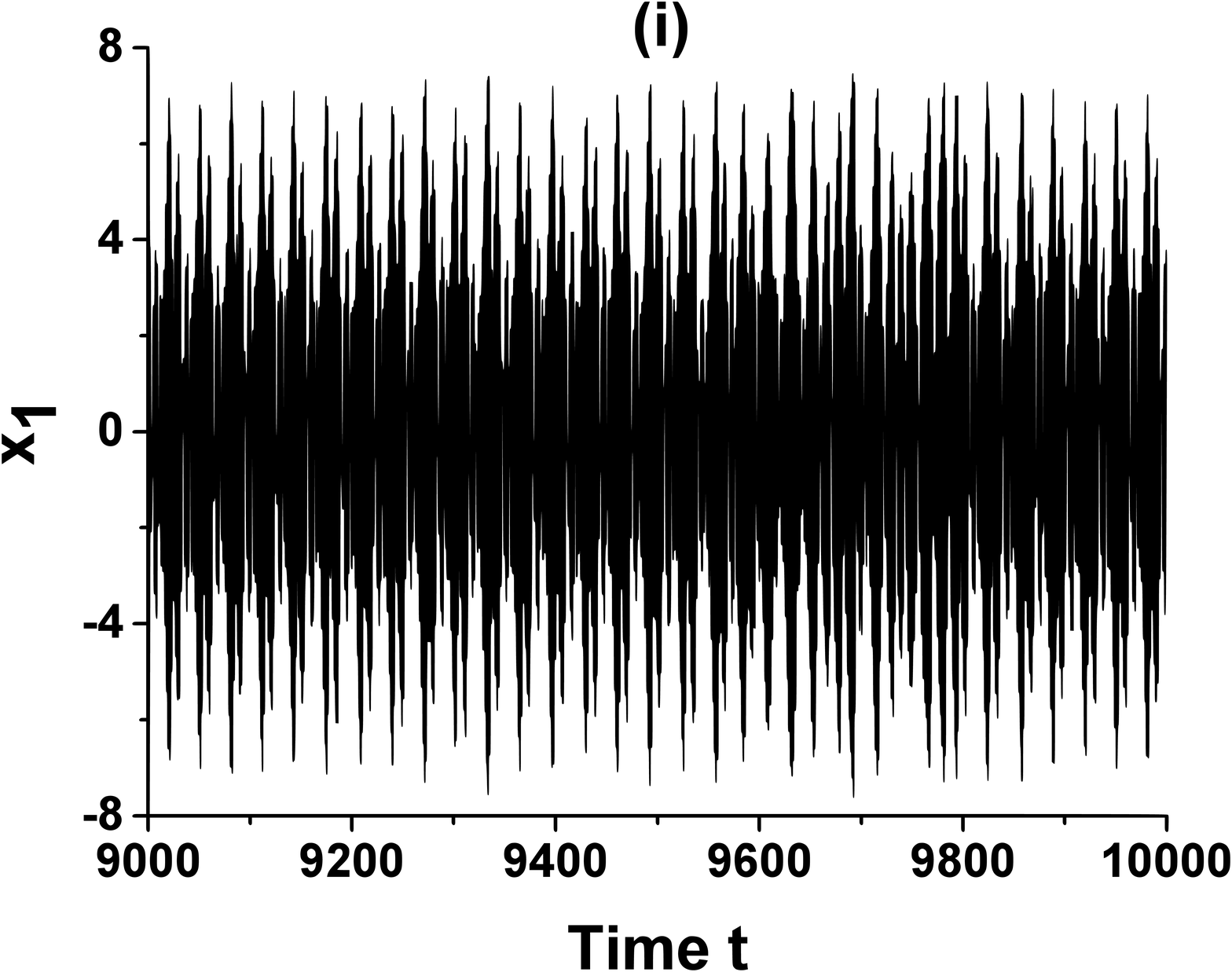} 
      \includegraphics[width=0.25\textwidth]{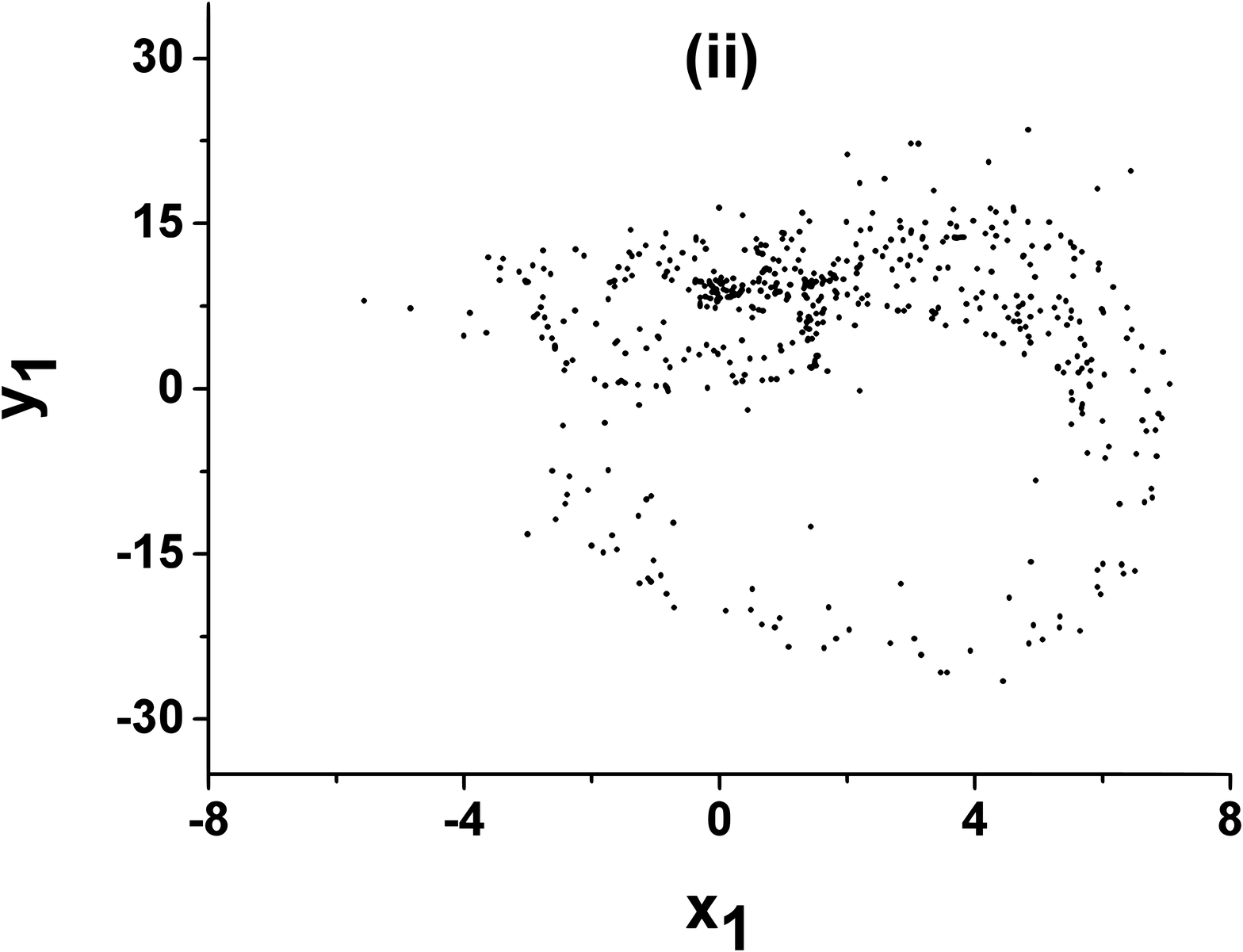} 
      \includegraphics[width=0.25\textwidth]{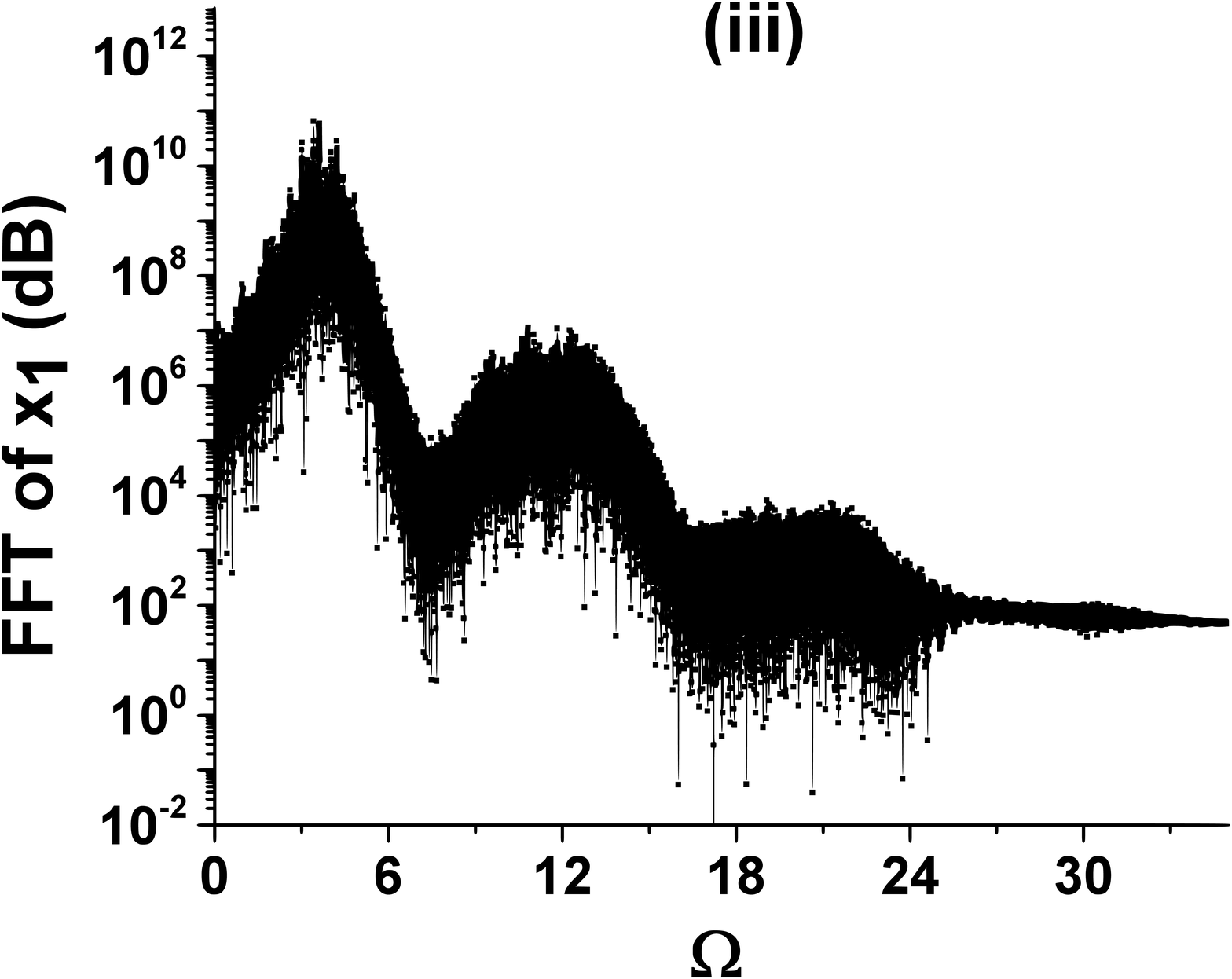}  \\
    \end{tabular}}
  \caption{(i) Time series, (ii) Poincar\'e sections, and (iii) power spectra for different values of the coupling strength (a) $\sigma=0.354$, (b) $\sigma=0.5$, (c) $\sigma=1.078$, (d) $\sigma=1.98$, and (d) $\sigma=3.25$. $\omega_0^2=-0.25$, $\delta=0.5$, $\alpha =0.4$.
    \label{fig2}}
\end{figure*}

In fig.~\ref{fig1} we observe the well-known scenario from a stable steady state to hyperchaos when the coupling strength $\sigma$ is increased. The details of dynamical regimes on this route are presented in fig.~\ref{fig2} with the time series, Poincar\'e sections, and power spectra. Starting from $\sigma=0$, first, a stable fixed point transforms into a limit cycle in the Hopf bifurcation when the coupling reaches $\sigma_{1} \approx 0.35$, where the largest Lyapunov exponent reaches zero (black line in Fig. \ref{fig2}(b)). This regime is illustrated in fig.~\ref{fig2}(a) and maintained within  a relatively small region of $0.35<\sigma<0.5$. Then, at $\sigma_2=0.5$ the limit cycle transforms into a quasiperiodic regime (2D torus) when the second frequency appears in the power spectrum (Fig. \ref{fig2}(b)). This quasiperiodic regime appears when the second largest Lyapunov exponent becomes zero and the third largest Lyapunov exponent approaches zero. When $\sigma$ is further increased, a 3D torus arises at $\sigma_3=1.0$ when the third largest Lyapunov exponent approaches zero. This regime is observed for $1.0<\sigma<1.75$ and characterized by a large number of frequencies in the power spectrum (fig.~\ref{fig2}(c)). At $\sigma_4=1.75$ the system becomes chaotic because the largest Lyapunov exponent becomes positive (fig.~\ref{fig2}(b)) and the power spectrum is wide (fig.~\ref{fig2}(d)). A further increase in the coupling strength leads to hiperchaos at $\sigma_5$ when two largest Lyapunov exponents become positive; this regime is illustrated in fig.~\ref{fig2}(e)). 

\begin{figure}[th!]
  \centerline{%
    \begin{tabular}{c} 
      \includegraphics[width=0.25\textwidth]{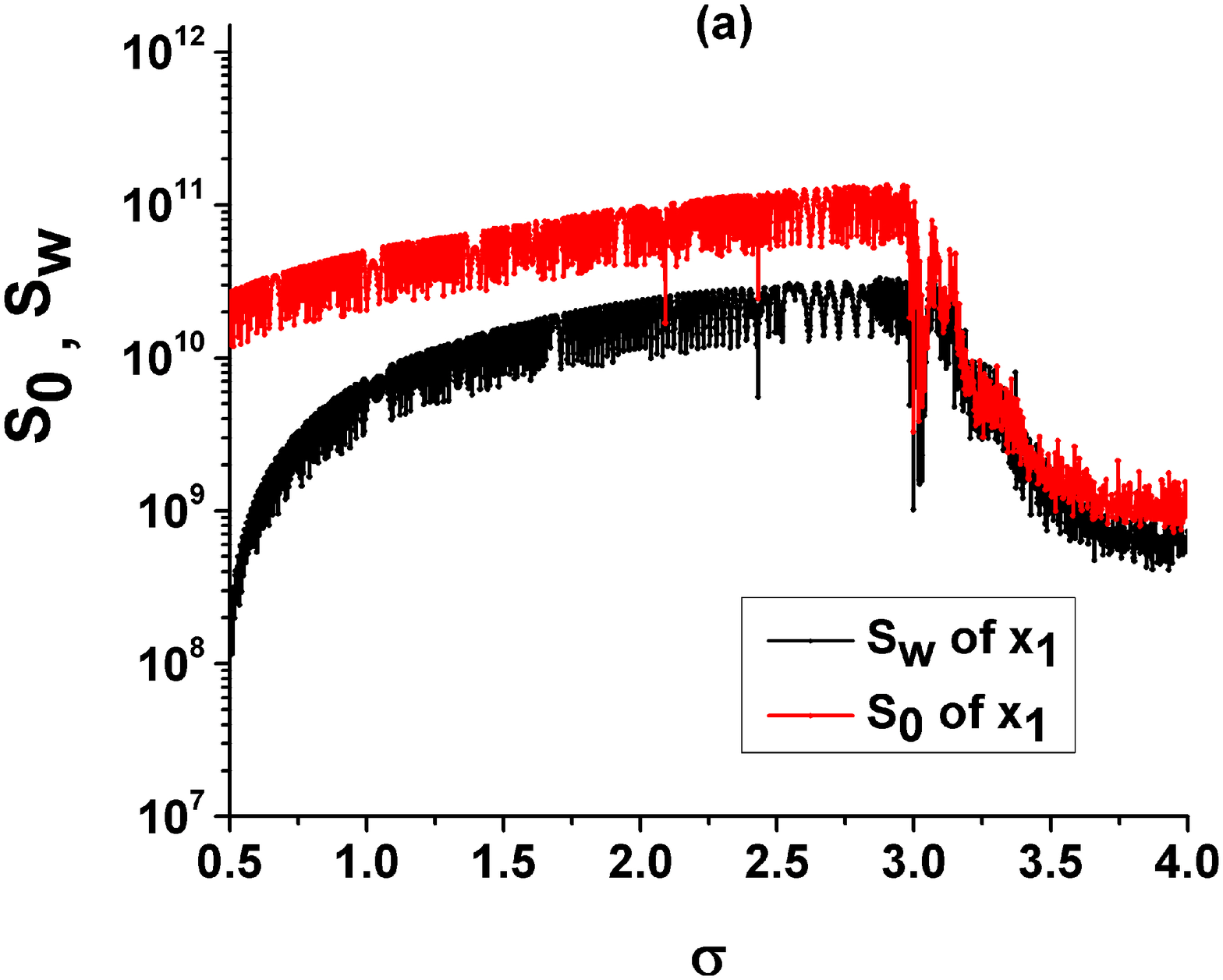}
      \includegraphics[width=0.25\textwidth]{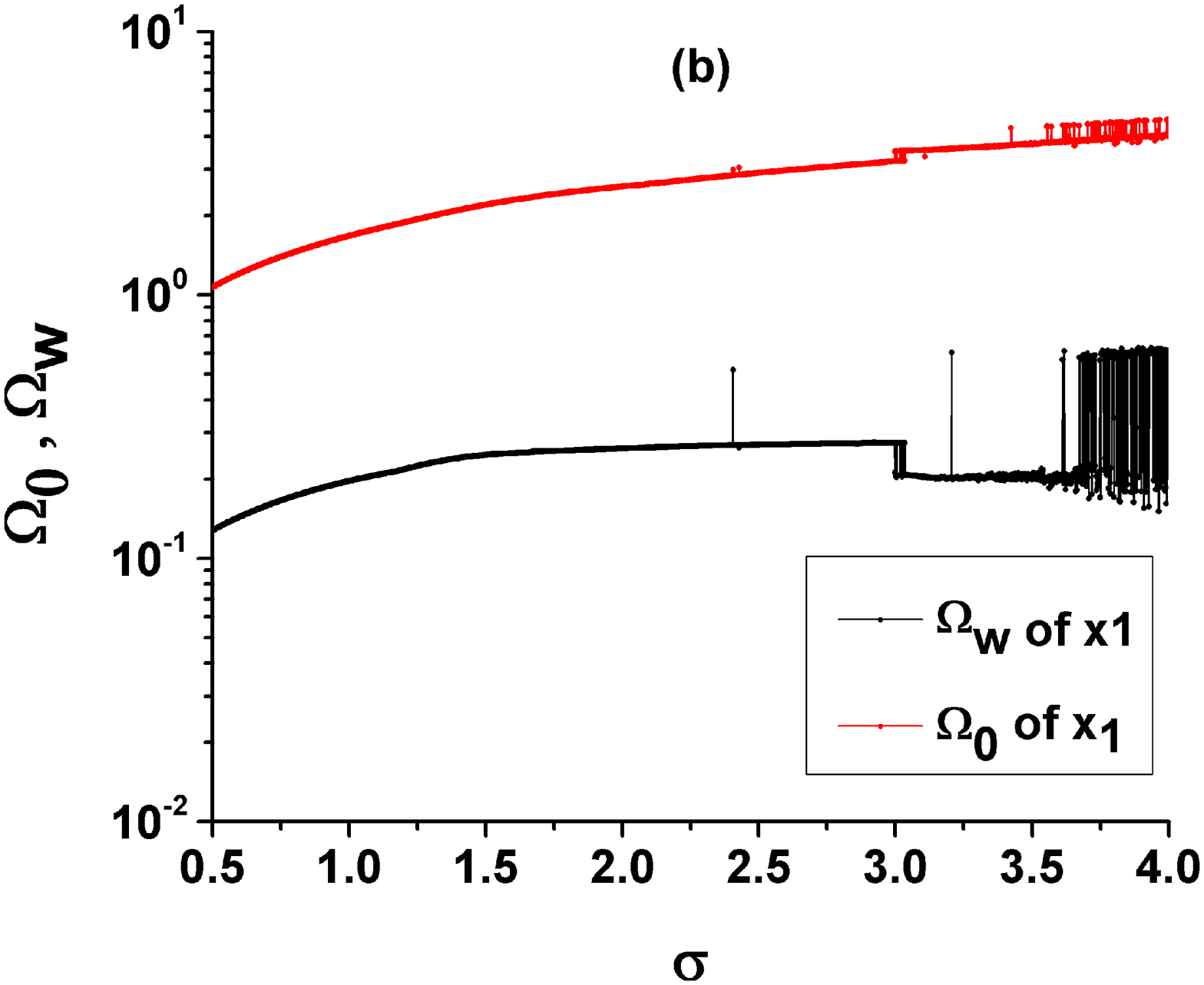}
    \end{tabular}}
  \caption{(a) Maximum ($S_0$) and rotating wave ($S_W$) spectral components and (b) their frequencies ($\Omega_0$ and $\Omega_W$) of $x_1$ versus coupling strength $\sigma$.  $\omega_0^2=-0.25$, $\delta=0.5$, $\alpha=0.4$. 
  \label{fig3}}
\end{figure}

Another interesting dynamical feature of the ring-coupled oscillators is the existence of the rotation wave in the quasiperiodic and chaotic regimes which consists in the constant phase difference between the envelope (second frequency) of quasiperiodic and chaotic oscillations of each oscillator. This fascinating synchronization state was first discovered in ring-coupled Chua oscillators~\cite{matias1997,marino1999} and then found in ring-coupled Lorenz~\cite{matias1998,deng2002,sanchez2006} and Duffing oscillators~\cite{perlikowski2010routes,borkowski2015experimental,borkowski2020stability}. 
In fig.~\ref{fig3}(a) we plot the maximum spectral component $S_{0}$ at the dominant frequency $\Omega_{0}$ (main oscillation frequency) and the rotating wave spectral power $S_{W}$ at the wave frequency $\Omega_{W}$ (envelope freqeuncy) as functions of the coupling strength. The latter is obtained from the power spectrum of the Hilbert transform and shown in fig.~\ref{fig3}(b). One can see from fig.~\ref{fig4}(a) that both spectral components increases as $\sigma$ is increased up to $\sigma=3$ and then the powers become decrease. This decrease happens because in this hyperchaotic regime the spectral energy is distributed over a wide frequency range, as seen in fig.~\ref{fig2}(e)(iii)). The wave frequency $S_{W}$ is almost independent of $\sigma$ in the chaotic and hyperchaotic regions, i.e. for $\sigma>1.75$, although for very large couplings ($\sigma>3$) it is not well determined. 

The underlying mechanism of the rotating wave stability is explained as the effect of an additional rotational degree of freedom and the symmetric structure of the coupling. Previous results associate the rotational wave with the space-time symmetry property of an invariant ring of identical oscillators under cyclic group~\cite{collins1994group,pazo2001transition}. Our results obtained with double-well Duffing oscillators confirm the previous findings. 

\section{Damping coefficient proportional to time ($\alpha(t)=t/4$)}
\begin{figure}[th!]
	\centering
        \includegraphics[width=0.24\textwidth]{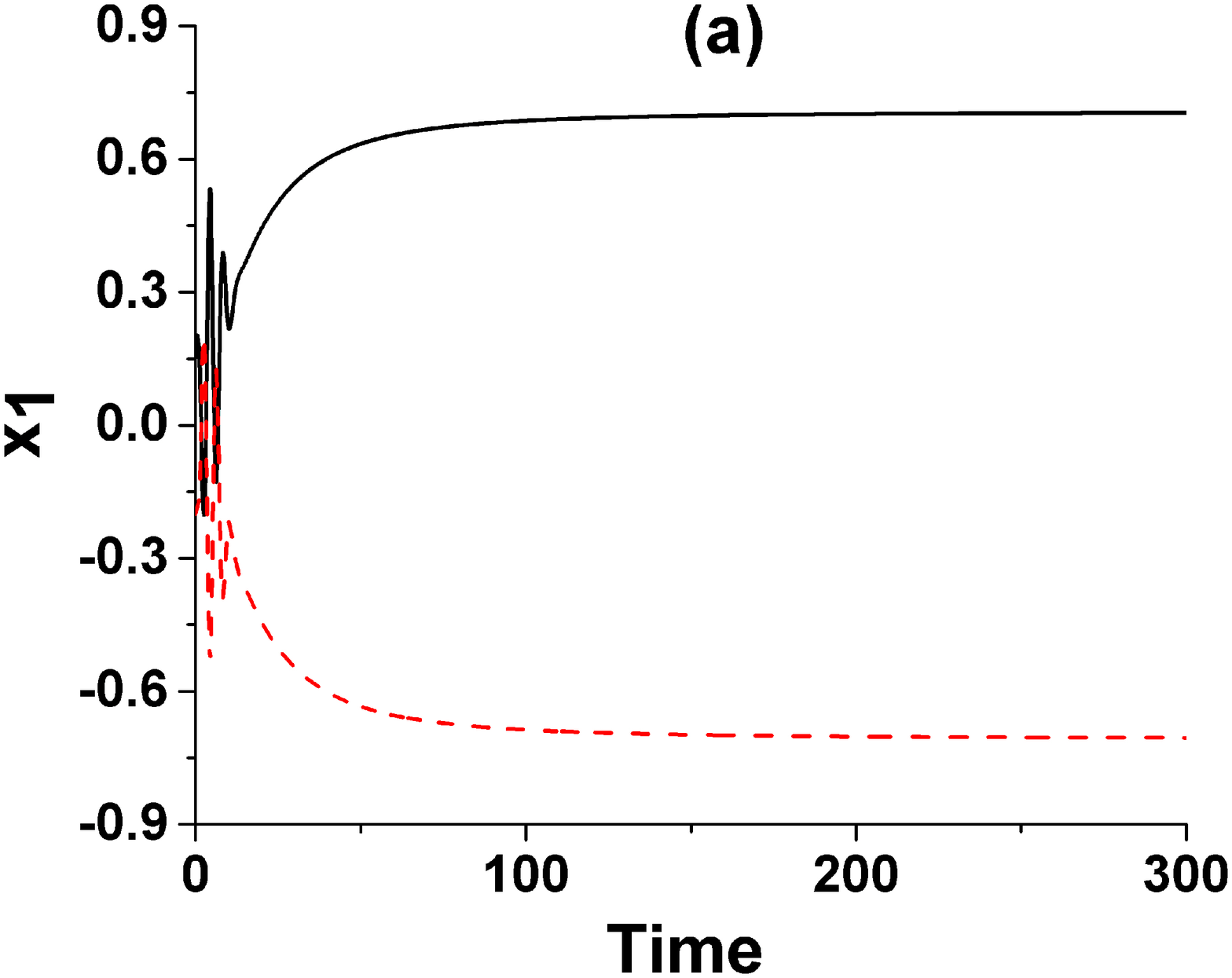}
        \includegraphics[width=0.24\textwidth]{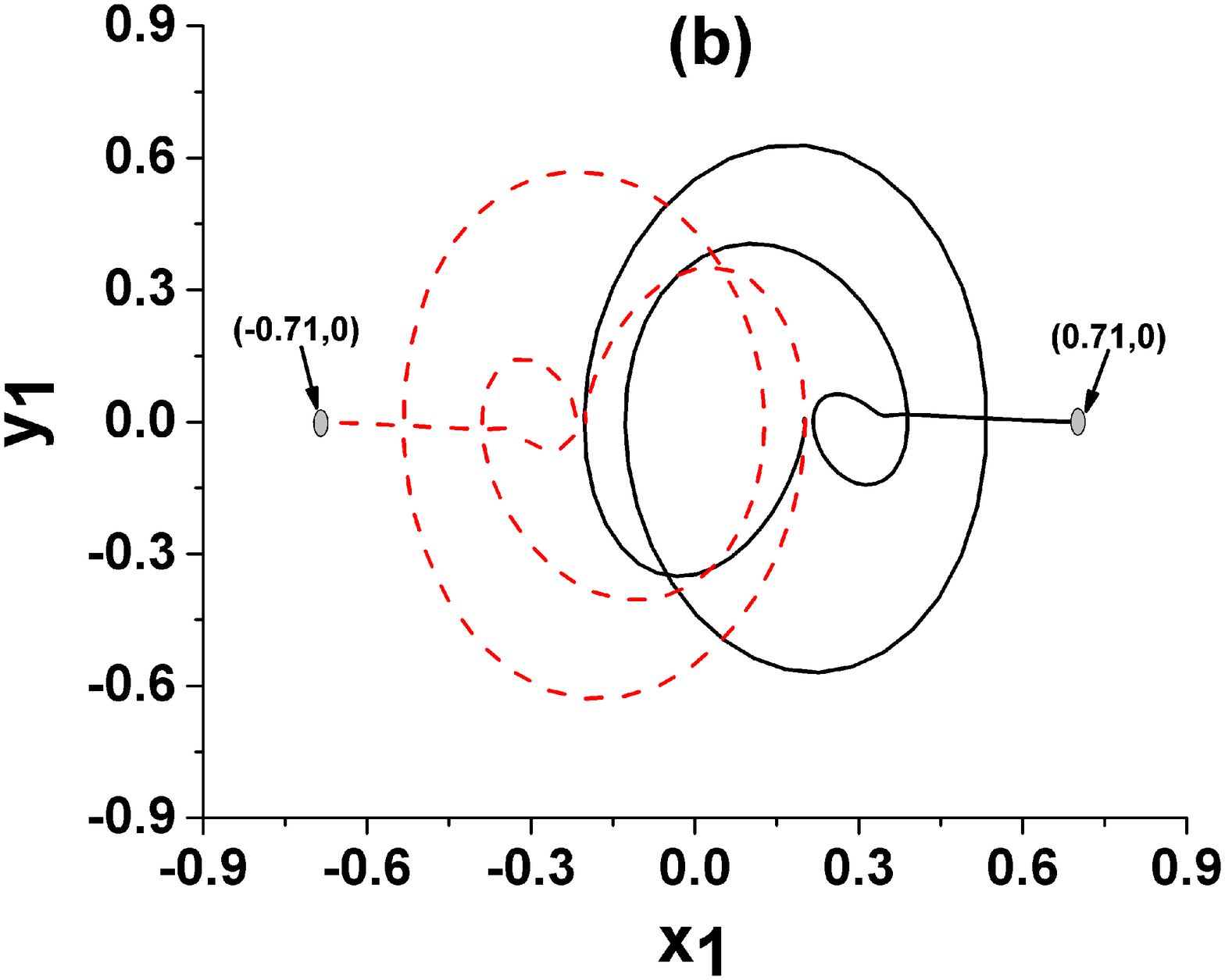}    
        \caption{(a) Time series and (b) phase portraits on the ($x_1$, $y_1$) plane for $\omega_0^2=-0.25$, $\delta=0.5$, and $\alpha(t)=t/4$.
          \label{fig4}}
\end{figure}

Let us now consider the second case, when the damping coefficient is proportional to time, specifically, $\alpha(t)=t/4$. In this case, the system is extremely dissipative and therefore the rotating wave does not appear. Instead, starting from different initial conditions the system tends to one of two stable equilibrium points ($x_{1}=\pm 0.71, y_1=0$) of the potential $V(x_1,t)$ (see eqs.~(\ref{POT1}) and (\ref{FP})), as illustrated with the time series and phase portraits in figs.~\ref{fig4}(a) and (b), respectively. 

\begin{figure}[th!]
  \centering
        \includegraphics[width=0.24\textwidth]{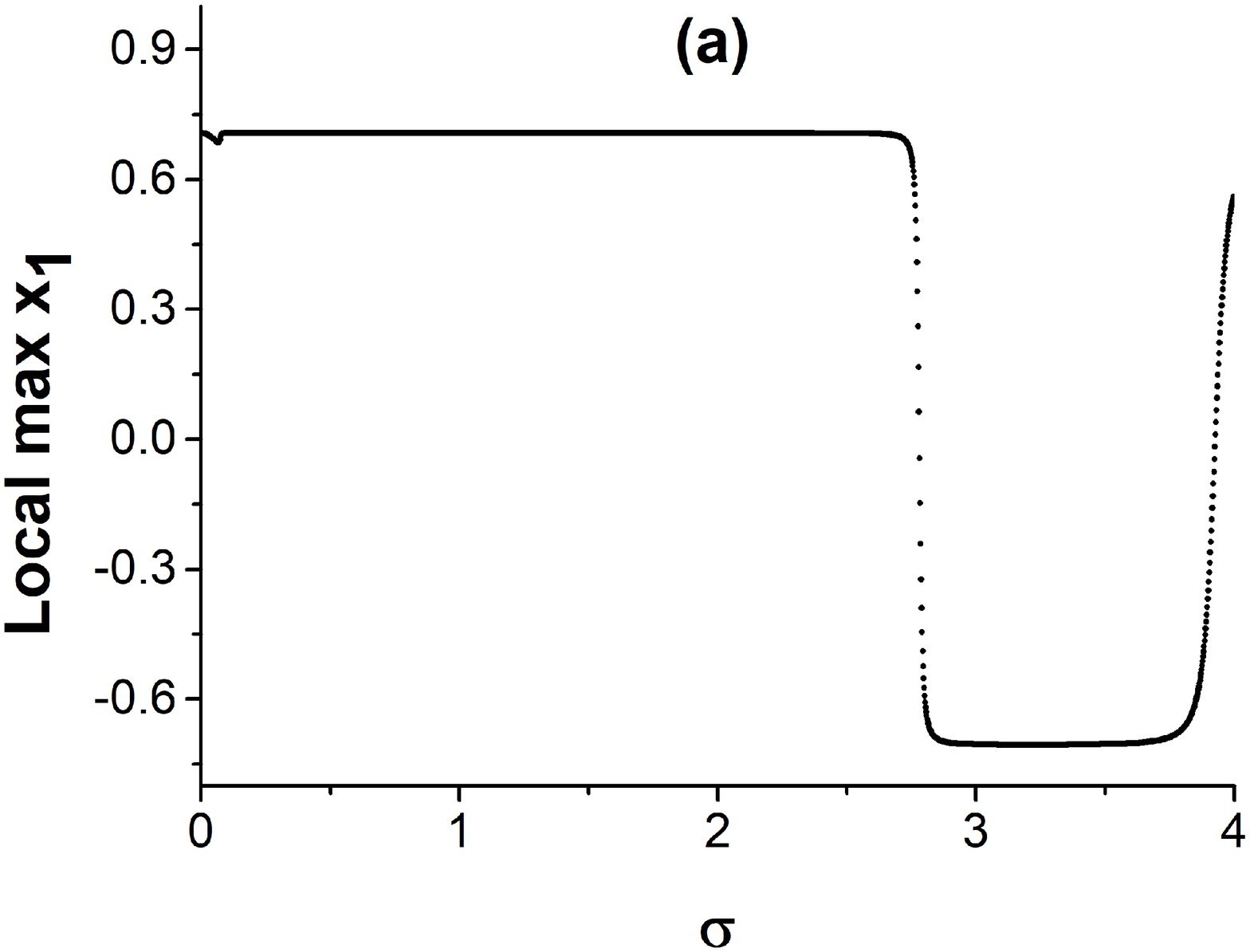}
        \includegraphics[width=0.24\textwidth]{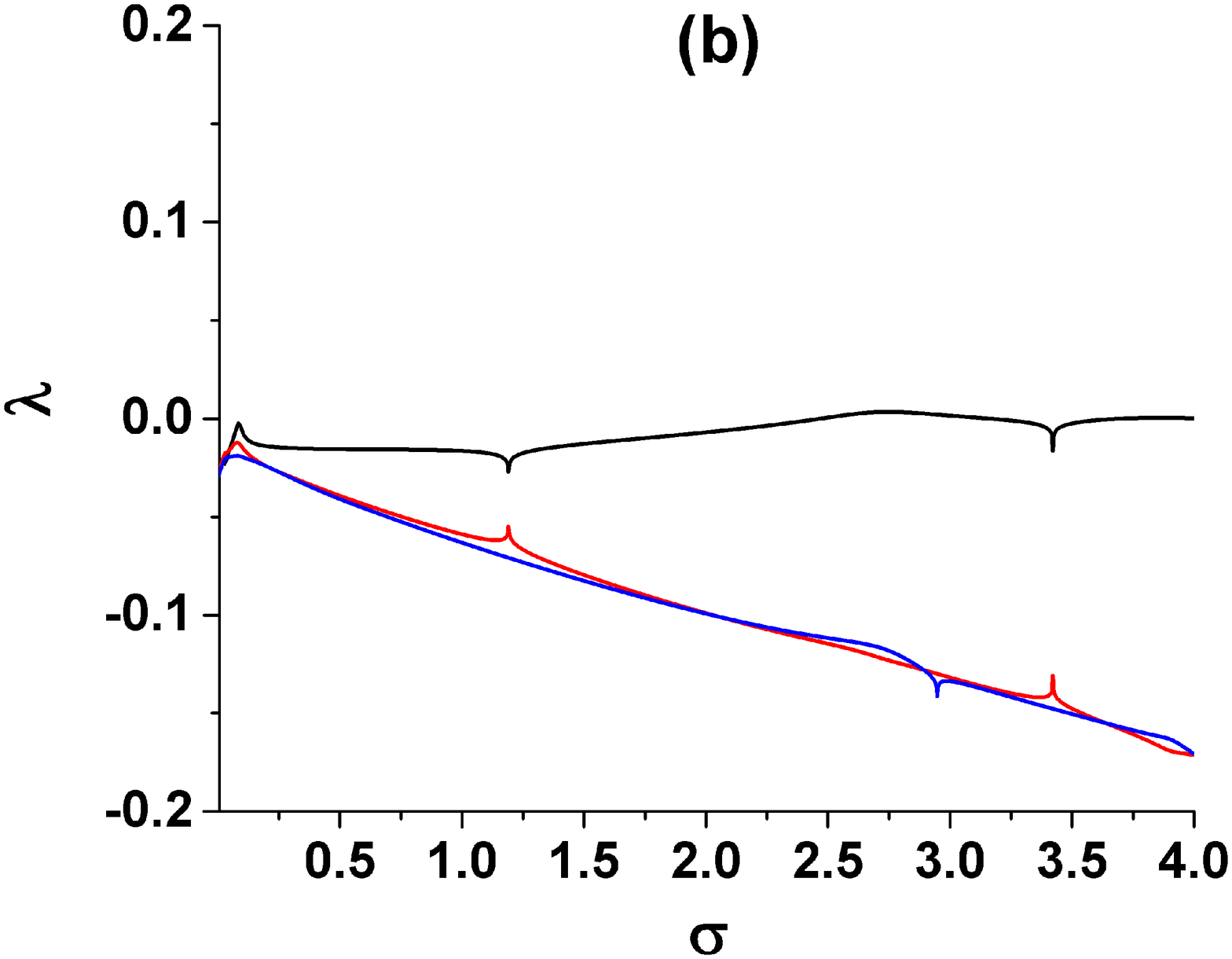} 
        \caption{(a) Bifurcation diagram of $x_1$ and (b) three largest Lyapunov exponents versus coupling strength for $\omega_0^2=-0.25$, $\delta=0.5$, and $\alpha(t)=t/4$.
          \label{fig5}}
\end{figure}

Figures~\ref{fig5}(a) and (b) show, respectively, the bifurcation diagram of oscillator $x_{1}$ and three largest Lyapunov exponent as functions of the coupling strength $\sigma$. One can see in fig.~\ref{fig5}(a) that the system switches to another coexisting fixed point although the initial conditions are fixed. This happens because for these values of the coupling parameters the initial conditions hit to the basin of attraction of another equilibrium. As seen from fig.~\ref{fig5}(b), all Lyapunov exponents are negative for any value of the coupling strength, and the system becomes more stable as $\sigma$ is increased. So, the high damping coefficient does not allow the rotating wave. 

\section{Damping coefficient inversely proportional to time ($\alpha(t)=1/t$)}
\begin{figure}[th!]
	 \centering
        \includegraphics[width=0.24\textwidth]{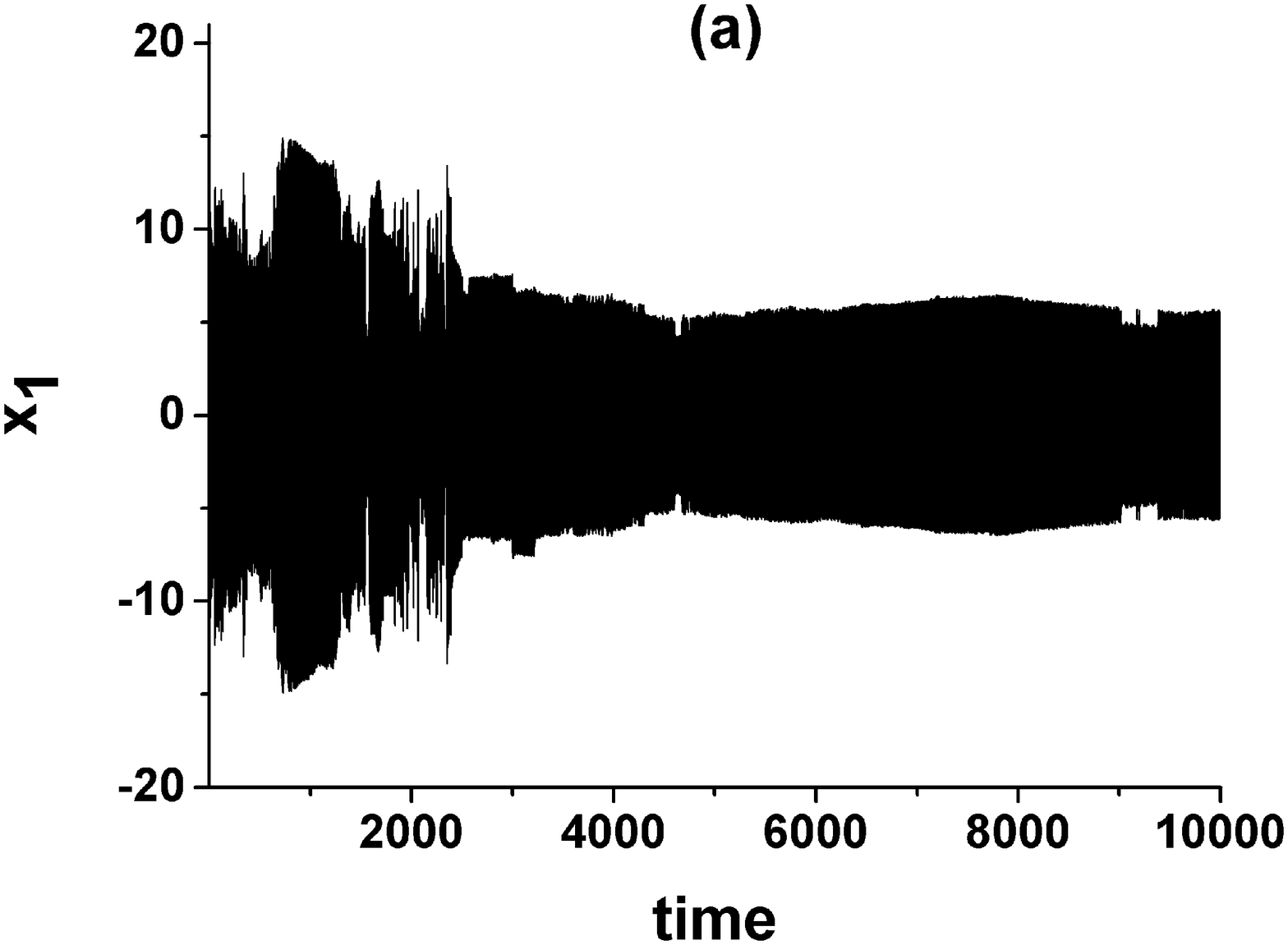}
        \includegraphics[width=0.24\textwidth]{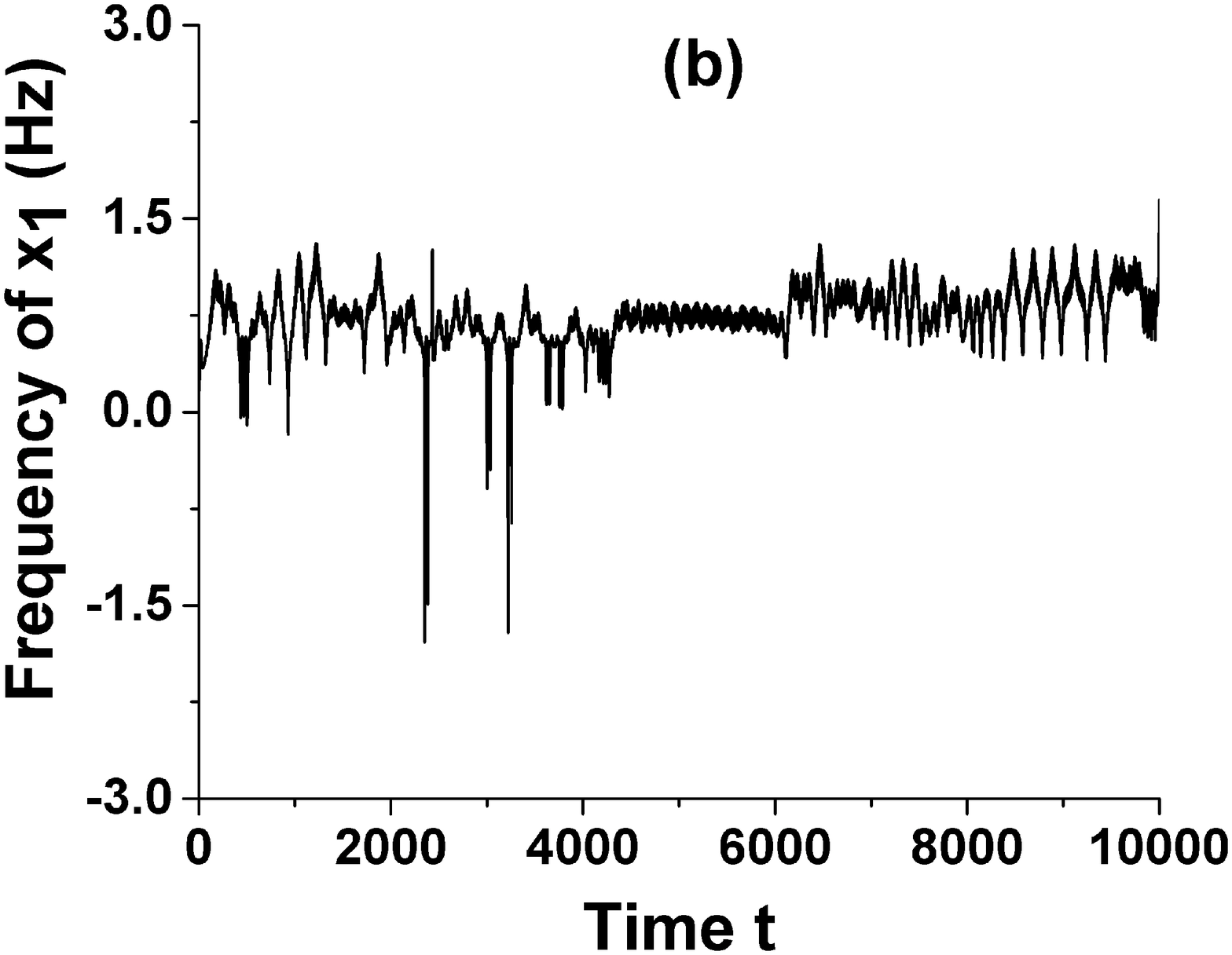} 
        \caption{(a) Time series of $x_1$ and (b) corresponding instantaneous frequency, representing transient hyperchaos. $\omega_0^2=-0.25$, $\delta=0.5$, $\sigma=3.25$, and $\alpha(t)=1/t$.
          \label{fig6}}
\end{figure}

Finally, we consider the case when the damping coefficient is inversely proportional to time, specifically, $\alpha(t)=1/t$, i.e., the damping tends to zero over time. The time series and corresponding instantaneous frequency of $x_1$ calculated using the Hilbert transform are shown in figs.~\ref{fig6}(a) and (b), respectively. One can see that such damping function results in the transient behavior followed by infinity. Moreover, in fig.~\ref{fig6}(b) one can observe continuous changes of the instantaneous frequency during all time evolution. This evidences the existence of a permanent hyperchaotic transitory. 

\begin{figure}[th!]
	 \centering
        \includegraphics[width=0.24\textwidth]{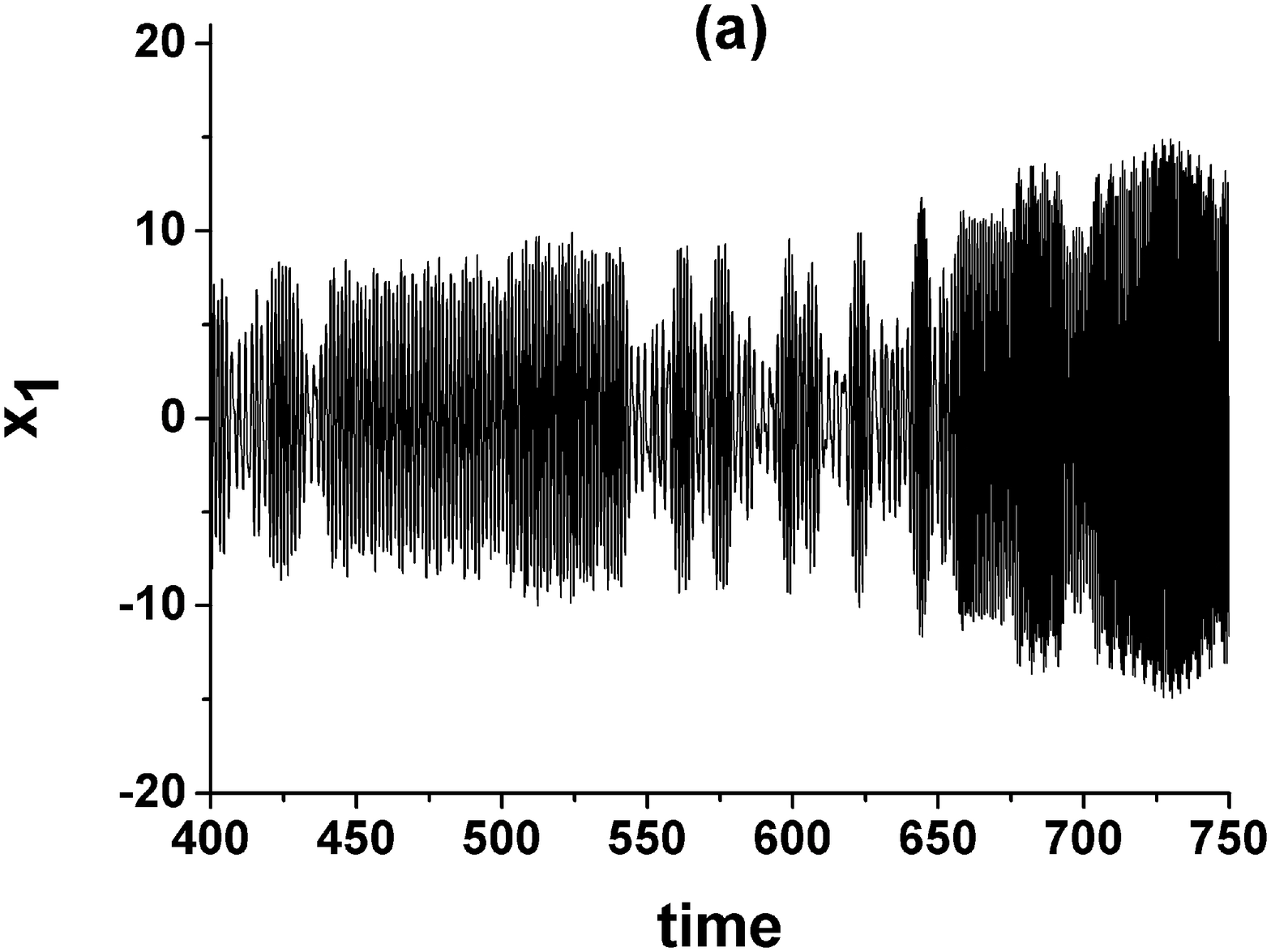} 
        \includegraphics[width=0.24\textwidth]{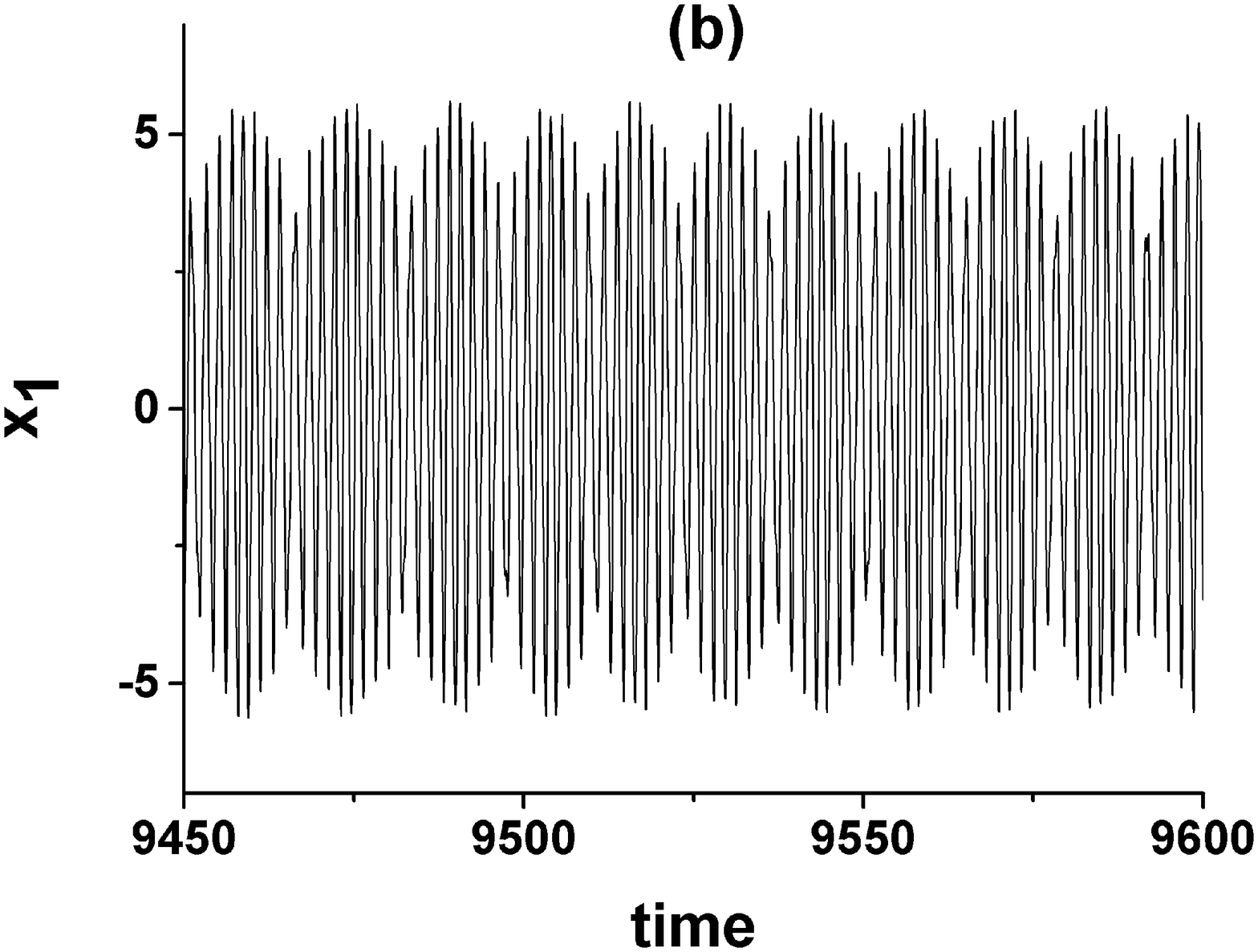}
        \caption{Time series in different periods of time representing (a) chaotic, (b) quasiperiodic of $x_1$ and (b) corresponding instantaneous frequency, representing transient hyperchaos. $\omega_0^2=-0.25$, $\delta=0.5$, $\sigma=3.25$, and $\alpha(t)=1/t$.
          \label{fig7}}
\end{figure}

The transient hyperchaos manifests itself as follows. The dissipative system eq.~(\ref{ODES1}) turns into conservative over time, causing hyperchaotic transient behavior, because a non-attracting chaotic set coexists with a chaotic attractor, i.e., there are two distinct forms of chaotic behavior. The trajectory proceeding from randomly chosen initial conditions looks chaotic for a sufficiently long period of time, during which it visits various chaotic and quasiperiodic states rather abruptly, as seen from the time series in fig.~\ref{fig7} corresponding to different time periods.

\begin{figure}[th!]
	 \centering
                \includegraphics[width=0.45\textwidth]{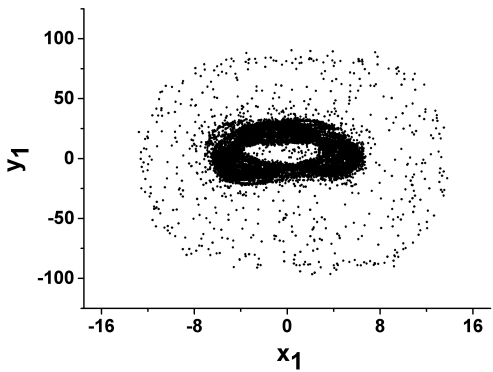} 
        \caption{Poincar\'e section on ($x_1$, $y_1$) plane representing transient toroidal hyperchaos for $\omega_0^2=-0.25$, $\delta=0.5$, $\sigma=3.25$, and $\alpha(t)=1/t$.
          \label{fig8}}
\end{figure}

Transient chaotic behavior appears in a conservative system where the phase-space volume is constant under time evolution~\cite{lai2011transient}. In the Poincar\'e section on ($x_1$, $y_1$) plane (fig.~\ref{fig8}) one can distinguish two regions. The region of many scattered points represents non-attracting chaotic set called chaotic saddle~\cite{sabarathinam2015transient} at the beginning of the transients (large amplitude oscillations at $t<2500$), whereas the central bagel of high density points corresponds to the resting part of the transients and represents toroidal hyperchaos.

\begin{figure}[th!]
	 \centering
        \includegraphics[width=0.24\textwidth]{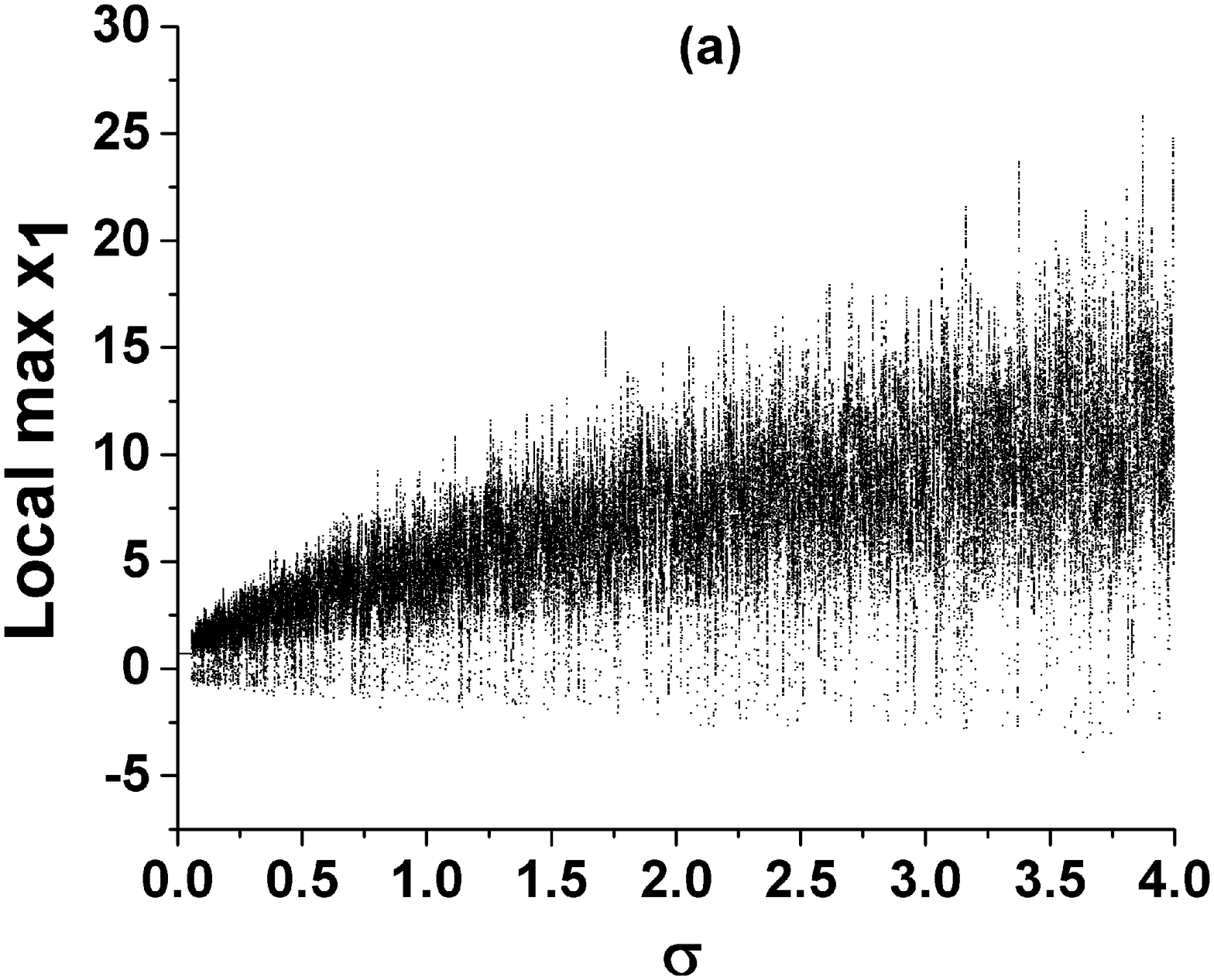}
        \includegraphics[width=0.24\textwidth]{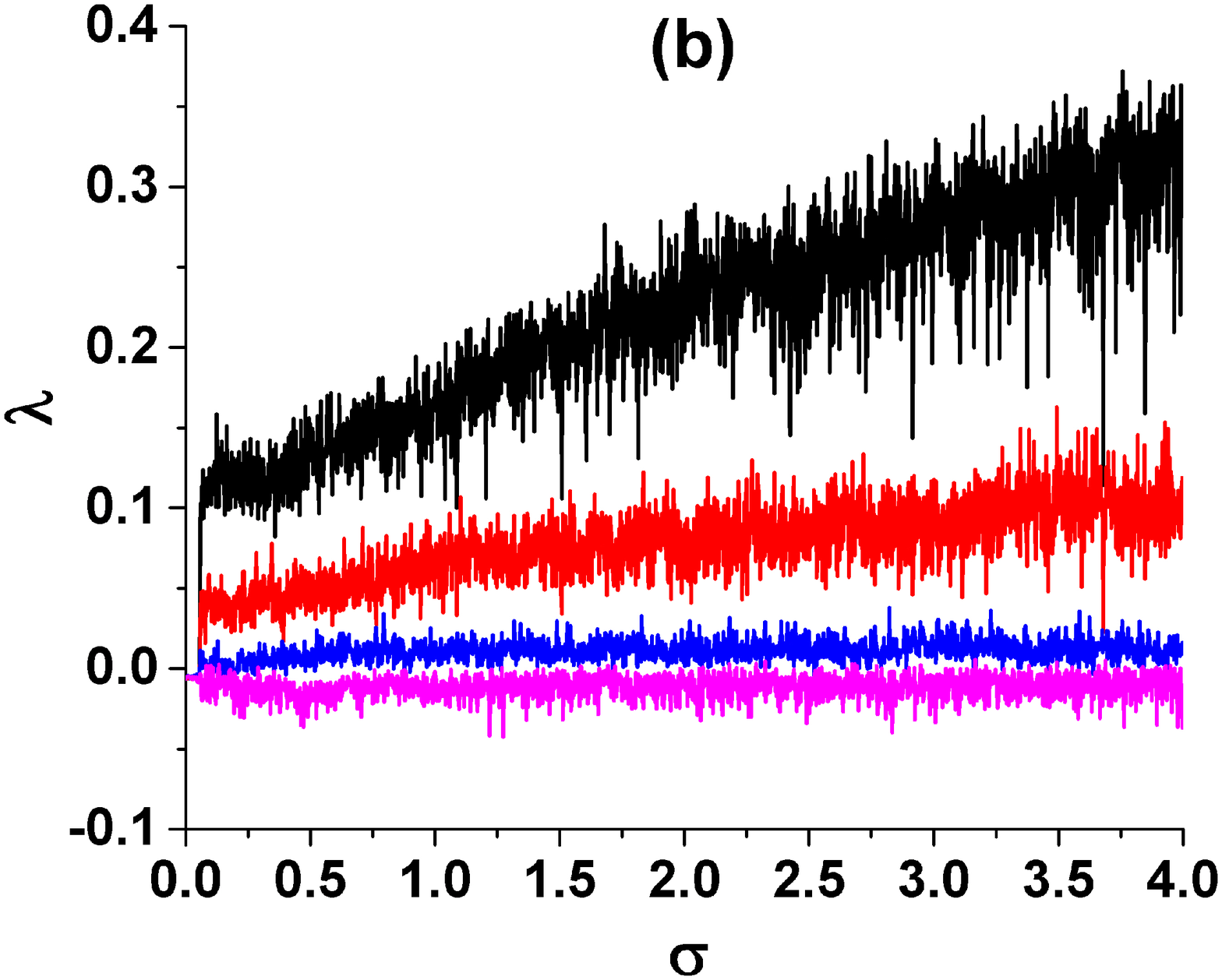}
        \caption{(a) Bifurcation diagram of $x_1$ and (b) largest Lyapunov exponents versus coupling strength for $\omega_0^2=-0.25$, $\delta=0.5$, $\sigma=3.25$, and $\alpha(t)=1/t$.
          \label{fig9}}
      \end{figure}

Figure~\ref{fig9}(a) and (b) show the bifurcation diagram of local maxima of $x_1$ and four largest Lyapunov exponents versus $\sigma$, respectively. One can see that when the coupling strength is increased, the system exhibits a fast transition to hyperchaos from a fixed point via a crisis bifurcation at a very small coupled strength $\sigma=0.056$. For larger $\sigma$, two largest Luyapunov exponents are always positive and the third exponent is zero. This means that the system is hyperchaotic.

%%%%%%%%%%%%%%%%%%%%%%%%%%%%%%%%%%%%%
   
\section{Conclusion}
In this Letter, we have studied the dynamics of a system of three double-well Duffing oscillators unidirectionally coupled in a ring. We have considered three cases depending on the damping coefficient: constant damping, damping proportional to time ($\alpha(t)=t/4$), and damping inversely proportional to time ($\alpha(t)=1/t$). The system dynamics have been analyzed using time series, Fourier and Hilbert transforms, Poincar\'e sections, and Lyapunov exponents. In the first case, we observed the route from a steady state to hyperchaos through a series of torus bifurcations, as the coupling strength is increased, as well as the existence of a rotating wave with a fixed phase between neighboring oscillators, which persists for quasiperiodic, chaotic and hyperchaotic regimes. These results are in a good agreement with previous studies of single-well Duffing oscillators in the same coupling configuration. Such similarity is explained by the fact that in the quasiperiodic and chaotic regimes, the system becomes monostable. 

In the second case, the system becomes highly dissipative and therefore does not generate a rotating wave. As a result, the system, is evolved to one of the stable steady states which depends on the initial conditions. During the transient behavior, the system can occasionally switch between two coexisting states. 

Finally, in the third case, the system becomes conservative when the time $t$ involves to large values, and then goes to infinity. The transient toroidal hyperchaotic behavior was observed. During transients, the system can visit  different unstable periodic orbits.  

Although this work was devoted to the study of the simple network motif of only three coupled oscillator, we suppose that similar dynamics may occur in larger oscillatory networks (see, e.g., \cite{milo2002network}), especially in a larger ring of unidirectionally coupled Duffing oscillators. This is a promising topic for future research.   

\acknowledgements
J. J. B. F. thanks the National Council for Science and Technology of Mexico (CONACYT) for the financial support granted through the scholarship number 924190. S. A. G. and A. N. P. acknowledge support from the Lobachevsky University Competitiveness Program in the frame of the 5-100 Russian Academic Excellence Project. 

%\section*{References}

%\bibliographystyle{elsarticle-num}
%\bibliography{Bibfile}

  \end{document}